\documentclass[12pt]{article}

\usepackage{psfrag,epsf}
\usepackage{enumerate}
\usepackage[pdftex]{graphicx}
\usepackage{amsfonts}
\usepackage{setspace}
\usepackage{amssymb}
\usepackage{amsthm}
\usepackage{amsmath}
\usepackage{mathtools}
\usepackage{mathrsfs}
\usepackage{caption}
\usepackage{subcaption}
\usepackage{color}
\definecolor{darkblue}{rgb}{0.0,0.0,0.55}
\RequirePackage[colorlinks,citecolor=darkblue,urlcolor=darkblue,linkcolor=darkblue]{hyperref} 
\usepackage{syntonly}
\usepackage{float}
\usepackage{booktabs}
\usepackage{bm}
\usepackage{algorithm}
\usepackage{algpseudocode}
\usepackage{bbold}
\usepackage[export]{adjustbox}
\usepackage{setspace}
\usepackage{titlesec}
\usepackage{lipsum}
\usepackage{natbib}
\usepackage{tabularx}
\usepackage{graphicx}
\usepackage{mwe}
\usepackage{url}
\usepackage{natbib}
\usepackage[flushleft]{threeparttable}
\allowdisplaybreaks

\usepackage[utf8]{inputenc}
\usepackage[english]{babel}

\newtheorem{theorem}{Theorem}[section]

\newtheorem{lemma}[theorem]{Lemma}

\newcommand{\RN}[1]{%
	\textup{\uppercase\expandafter{\romannumeral#1}}%
}

\newcommand{\blind}{0}

\begin{document}

\def\spacingset#1{\renewcommand{\baselinestretch}%
	{#1}\small\normalsize} \spacingset{1}
	

\if0\blind
{
	\title{\vspace{-1cm} \bf Joint Mean--Covariance Estimation via the Horseshoe with an Application in Genomic Data Analysis}
	\author{Yunfan Li \\
		Department of Statistics, Purdue University\\
		\\
		Jyotishka Datta\\
		Department of Mathematical Sciences, University of Arkansas\\
		\\
		Bruce A. Craig \\
		Department of Statistics, Purdue University\\
		\\
		Anindya Bhadra \\
		Department of Statistics, Purdue University
		}
		\date{}
		\maketitle
		\thispagestyle{empty}
} \fi

\if1\blind
{
	\bigskip
	\bigskip
	\bigskip
	\begin{center}
		{\LARGE\bf Joint Mean--Covariance Estimation via the Horseshoe with an Application in Genomic Data Analysis}
	\end{center}
	\medskip
} \fi

\vspace{-0.5cm}
\begin{abstract}
	\noindent Seemingly unrelated regression is a natural framework for regressing multiple correlated responses on multiple predictors. The model is very flexible, with multiple linear regression and covariance selection models being special cases. However, its practical deployment in genomic data analysis under a Bayesian framework is limited due to both statistical and computational challenges. The statistical challenge is that one needs to infer both the mean vector and the inverse covariance matrix, a problem inherently more complex than separately estimating each. The computational challenge is due to the dimensionality of the parameter space that routinely exceeds the sample size. We propose the use of horseshoe priors on both the mean vector and the inverse covariance matrix. This prior has demonstrated excellent performance when estimating a mean vector or inverse covariance matrix separately. The current work shows these advantages are also present when addressing both simultaneously. A full Bayesian treatment is proposed, with a sampling algorithm that is linear in the number of predictors. MATLAB code implementing the algorithm is freely available from github at \url{https://github.com/liyf1988/HS\_GHS}. Extensive performance comparisons are provided with both frequentist and Bayesian alternatives, and both estimation and prediction performances are verified on a genomic data set.
\end{abstract}

\noindent%
\vfill

\clearpage\pagebreak\newpage
\pagenumbering{arabic}
\newlength{\gnat}
\setlength{\gnat}{26pt}
\baselineskip=\gnat

\section{Introduction}
\label{sec:intro}

Multivariate regression is ubiquitous in quantitative disciplines as diverse as finance and chemometrics. In recent years, multivariate regression has also been used in genomics, most notably in expression quantitative trait loci (eQTL) analysis, where the high dimensionality of the data necessitates the use of regularization methods and poses both theoretical and computational challenges.  An eQTL analysis typically involves simultaneously regressing the expression levels of multiple genes on multiple markers or regions of genetic variation. Early studies have shown that each gene expression level is expected to be affected by only a few genomic regions \citep{schadt2003genetics, brem2005landscape} so that the regression coefficients in this application are expected to be sparse. In addition, the expression levels of multiple genes have been shown to possess a sparse network structure that encodes conditional independence relationships \citep{leclerc2008survival}, which, in the case of a Gaussian model, are encoded by the off-diagonal zeros in the inverse covariance matrix. Therefore, an eQTL analysis, if formulated as a multiple predictor--multiple responses regression problem, presents with non i.i.d. error terms. In high dimensions, this necessitates regularized estimates of both the regression coefficients and the error inverse covariance matrix.

A natural question then is: what is there to be gained by treating all responses jointly rather than separately regressing each response (e.g., gene expressions) on the set of covariates (e.g., single nucleotide polymorphisms), possibly adjusting for multiplicity in the responses? In multivariate regression problems with correlated error terms, early works by \cite{zellner1962efficient} established that joint estimation of regression coefficients improves efficiency. \cite{zellner1962efficient} went on to propose the seemingly unrelated regression (SUR) framework where the error correlation structure in multiple responses is leveraged to achieve a more efficient estimator of the regression coefficients compared to separate least squares estimators. \cite{holmes2002accounting} adopted the SUR framework in Bayesian regressions. However, these early methods in the SUR framework considered a relatively modest dimension of the responses, and did not encourage sparse estimates of either the regression coefficients or the error inverse covariance matrix. Therefore, these methods can not be applied directly to analyze modern genomic data. Much more recently, both Bayesian and frequentist approaches that encourage sparsity have started to attract considerable attention in the eQTL analysis problem formulated in an SUR framework \citep[e.g.,][]{bhadra2013joint,yin2011sparse, cai2012covariate,touloumis2016hdtd, banterle2018sparse}. Precise descriptions of some of these competing approaches and understanding their strengths and limitations require some mathematical formalism. This is reserved for Section~\ref{subsec:related}.

In this article, we propose a \emph{fully Bayesian method} for high-dimensional SUR problems with an algorithm for efficient exploration of the posterior, which is ideally suited for application in eQTL analysis. We impose the horseshoe prior \citep{carvalho2010horseshoe} on the regression coefficients, and the graphical horseshoe prior \citep{li2017graphical} on the precision matrix. In univariate normal regressions, the horseshoe prior has been shown to possess many attractive theoretical properties, including improved Kullback--Leibler risk bounds \citep{carvalho2010horseshoe}, asymptotic optimality in testing under $0$--$1$ loss \citep{datta2013asymptotic}, minimaxity in estimation under the $\ell_2$ loss \citep{van2014horseshoe}, and improved risk properties in linear regression \citep{bhadra2016prediction}. The graphical horseshoe prior inherits the properties of improved Kullback--Leibler risk bounds, and nearly unbiased estimates, when applied to precision matrix estimation \citep{li2017graphical}. 

The beneficial theoretical and computational properties of the horseshoe (HS) and graphical horseshoe (GHS) are combined in our proposed method, resulting in a prior that we term HS-GHS. The proposed method is fully Bayesian, so that the posterior distribution can be used for uncertainty quantification, which in the case of horseshoe is known to give good frequentist coverage \citep{van2017uncertainty}. For estimation, we derive a full Gibbs sampler, inheriting the benefits of automatic tuning and no rejection that come with it. The complexity of the proposed algorithm is linear in the number of covariates and cubic in the number of responses. To our knowledge, this is the first fully Bayesian algorithm with a linear scaling in the number of covariates that allows \emph{arbitrary sparsity patterns} in both the regression coefficients and the error precision matrix.  This is at a contrast with existing Bayesian methods that require far more restrictive assumptions on the nature of associations. For example, \citet{bhadra2013joint} require that either a predictor is important to all the responses, or to none of them.  The proposed method is also at a contrast with approaches that require special structures on the conditional independence relationships. For example, both \citet{bhadra2013joint} and  \citet{banterle2018sparse} require that the graphical model underlying the inverse covariance matrix is \emph{decomposable}. Such assumptions are typically made for computational convenience, rather than any inherent biological motivations, and the current work delineates a path forward by dispensing with them. In addition to these methodological innovations, the performance of the proposed method is compared with several competing approaches in a yeast eQTL data set and superior performances in both estimation and prediction are demonstrated.

The remainder of this article is organized as follows. Section~\ref{subsec:related} formulates the problem and describes previous works in high-dimensional SUR settings, with brief descriptions of their respective strengths and limitations. Section~\ref{sec:model} describes our proposed HS-GHS model and estimation algorithm. Section~\ref{sec:KL} discusses theoretical properties in terms of Kullback--Leibler divergence between the true sampling density and the marginal density under the HS-GHS prior. In Section~\ref{sec:sim}, we evaluate the performance of our model in four simulation settings and compare them with results by competing approaches. Section~\ref{sec:yeast} describes an application in an eQTL analysis problem. We conclude by identifying some possible directions for future investigations.

\section{Problem Formulation and Related Works in High-Dimensional Joint Mean--Covariance Modeling}
\label{subsec:related}

Consider regressing responses $Y_{n \times q}$ on predictors $X_{n \times p}$, where $n$ is the sample size, $p$ is the number of features, and $q$ is the number of possibly correlated outcomes. A reasonable parametric linear model is of the form $Y_{n \times q} = X_{n \times p}B_{p\times q} + E_{n\times q}$, where $E \sim \mathrm{MN}_{n\times q} (0, I_n, \Omega_{q\times q}^{-1})$ denotes a matrix normal random variate \citep{dawid1981some} with the property that $vec(E') \sim \mathrm{N}_{nq} (0, I_n \otimes \Omega_{q\times q}^{-1})$, a multivariate normal, where $vec(A)$ converts a matrix $A$ into a column vector by stacking the columns of $A$, the identity matrix of size $n$ is denoted by $I_n$, and $\otimes$ denotes the Kronecker product. Thus, this formulation indicates the $n$ outcome vectors of length $q$ are assumed uncorrelated, but within each outcome vector, the $q$ responses share a network structure, which is reasonable for an eQTL analysis. The problem is then to estimate $B_{p \times q}$ and $\Omega_{q \times q}$. We drop the subscripts denoting the dimensions henceforth when there is no ambiguity. Here $\Omega$ is also referred to as the precision matrix of the matrix variate normal, and off-diagonal zeros in it encodes a conditional independence structure across the $q$ responses, after accounting for the covariates. Of course, a consequence of the model is that one has conditionally independent (but not i.i.d.) observations of the form $Y_i \sim \textrm{N}(X_i B, \Omega^{-1})$, for $i=1,\ldots, n$. The negative log likelihood function under this model, up to a constant, is
\begin{equation*}
	l(B,\Omega) = \mathrm{tr}\{n^{-1}(Y-XB)'(Y-XB)\Omega\}-\mathrm{log}|\Omega|.
\end{equation*}
\noindent The maximum likelihood estimator for $B$ is simply $\hat{B}^{OLS} = (X'X)^{-1}X'Y$, which does not exist when $p>n$. In addition, increasing $|\Omega|$ easily results in an unbounded likelihood function. Therefore, many methods seek to regularize both $B$ and $\Omega$ for well-behaved estimates.

One of the earliest works in high dimensions is the multivariate regression with covariance estimation or the MRCE method \citep{rothman2010sparse}, which adds independent $\ell_1$ penalties to $B$ and $\Omega$, so the objective function is
\begin{equation*}
	(\hat{B}_{MRCE},\hat{\Omega}_{MRCE}) = \underset{(B,\Omega)}{\mathrm{argmin}} \Big \{l(B,\Omega)+\lambda_1 \Sigma_{k \neq l}|\omega_{kl}| + \lambda_2 \Sigma_{j=1}^{pq} |\beta_j| \Big \},
\end{equation*}

\noindent where $\omega_{kl}$ are the elements of $\Omega$, $\beta_j$ are the elements of vectorized $B'$, and $\lambda_1, \lambda_2 >0$ are tuning parameters. A coordinate descent algorithm is developed that iteratively solves a lasso and a graphical lasso problem to update $\hat{B}_{MRCE}$ and $\hat{\Omega}_{MRCE}$, respectively.

\cite{cai2012covariate} developed the covariate-adjusted precision matrix estimation or CAPME procedure taking a two-stage approach and using a multivariate extension of the Dantzig selector of \citet{candes2007dantzig}. Let $\bar{y}=n^{-1}\Sigma_{i=1}^n y_i$, $\bar{x}=n^{-1}\Sigma_{i=1}^n x_i$, $S_{xy}=n^{-1}\Sigma_{i=1}^n (y_i-\bar{y})(x_i-\bar{x})'$ and $S_{xx}=n^{-1}\Sigma_{i=1}^n (x_i-\bar{x})(x_i-\bar{x})'$. The estimate of $B$ in CAPME solves the optimization problem
\begin{equation*}
	\hat{B}_{CAPME} = \underset{B}{\mathrm{argmin}} \Big\{ |B|_1: |S_{xy}-B S_{xx}|_{\infty} \leq \lambda_n \Big\},
\end{equation*}

\noindent where $\lambda_n$ is a tuning parameter, $|A|_1$ defines the element-wise $\ell_1$ norm of matrix $A$, and $|A|_{\infty}$ defines the element-wise $\ell_{\infty}$ norm of $A$. This is equivalent to a Dantzig selector applied on the coefficients in a column-wise way. After inserting the estimator $\hat{B}_{CAPME}$ to obtain $S_{yy}=n^{-1} \Sigma_{i=1}^n (y_i-\hat{B}x_i)(y_i-\hat{B}x_i)'$, one estimates $\Omega$ by the solution to the optimization problem
\begin{equation*}
	\hat{\Omega}_{CAPME} = \underset{\Omega}{\mathrm{argmin}} \Big\{ |\Omega|_1: |I_p-S_{yy}\Omega|_{\infty} \leq \tau_n \Big\},
\end{equation*}

\noindent where $\tau_n$ is a tuning parameter. The final estimator of $\Omega$ needs to be symmetrized since no symmetry condition on $\Omega$ is imposed.

Critiques of the lasso shrinkage include that the lasso estimate is not tail robust \citep{carvalho2010horseshoe}, and at least empirically, the Dantzig selector rarely outperforms the lasso in simulations and in genomic data sets \citep{meinshausen2007discussion,zheng2011experimental}, indicating these problems might be inherited by MRCE and CAPME, respectively.

Bayesian approaches seek to implement  regularization through the choice of prior, with the ultimate goal being probabilistic uncertainty quantification using the full posterior. \cite{deshpande2017simultaneous} put spike-and-slab lasso priors on the elements of $B$. That is, $\beta_{kj}, k=1,\ldots, p; j=1,\ldots, q$ is drawn \textit{a priori} from either a `spike' Laplace distribution with a sharp peak around zero, or a `slab' Laplace distribution that is relatively flatter. A binary variable indicates whether a coefficient is drawn from the spike or the slab distribution. Such an element-wise prior on $\beta_{kj}$ is
\begin{equation*}
	\pi(\beta_{kj}|\gamma_{kj}) \propto (\lambda_1 e^{-\lambda_1|\beta_{kj}|})^{\gamma_{kj}}(\lambda_0 e^{-\lambda_0|\beta_{kj}|})^{1-\gamma_{kj}},
\end{equation*}

\noindent where $\lambda_1$ and $\lambda_0$ are the parameters for the spike and slab Laplace distributions, and the binary indicator $\gamma_{kj}$ follows a priori a Bernoulli distribution with parameter $\theta$, with a beta hyperprior distribution on $\theta$ with parameters $a_{\theta}$ and $b_{\theta}$. Similarly, spike-and-slab lasso priors are put on elements $\omega_{lm}$ in $\Omega$ as well. An Expectation/Conditional Maximization (ECM) algorithm is derived for this model to obtain the posterior mode. The hyper-parameters $(\lambda_1, \lambda_0, a_{\theta}, b_{\theta})$ for $B$, and the corresponding four hyper-parameters for $\Omega$, need to be specified in order to apply the ECM algorithm. In \cite{deshpande2017simultaneous}, the Laplace distribution hyper-parameters are chosen by the trajectories of individual parameter estimates given a path of hyper-parameters, and the beta hyper-parameters are set at predefined levels. The method does not provide samples from the full posterior.

\citet{bhadra2013joint} also consider a spike-and-slab prior on $B$ but place Bernoulli indicators in a different way. Their priors on $B$ and $\Omega^{-1}$ are
\begin{align*}
	B \mid \gamma,\Omega^{-1} &\sim \mathrm{MN}(0,cI_{p_\gamma},\Omega^{-1}), \\
	\Omega^{-1}\mid G &\sim \mathrm{HIW}_G(b,d I_q),
\end{align*}

\noindent where $b,c,d$ are fixed, positive hyper-parameters and HIW denotes the hyper-inverse Wishart distribution \citep{dawid1993hyper}. The vector of indicators $\gamma$ selects entire rows of coefficients, depending on whether $\gamma_i = 1;\;i=1,\ldots,p$. Similarly, the indicator $G$ has length $q(q-1)/2$, and selects the off-diagonal elements in the precision matrix. Here $p_\gamma=\sum_{i=1}^{p} \gamma_i$. Elements in $\gamma$ and $G$ are independently distributed Bernoulli random variables, with hyper-parameters $\omega_{\gamma}$ and $\omega_G$, respectively. The model allows $B$ and $\Omega$ to be analytically integrated out to achieve fast Markov chain Monte Carlo (MCMC) sampling, at the expense of a somewhat restrictive assumption that a variable is selected as relevant to all of the $q$ responses or to none of them.

Thus, it appears only a few of Bayesian shrinkage rules have been applied to joint mean and inverse covariance estimation in SUR models, and there is no fully Bayesian method that efficiently solves this problem under the assumption of arbitrary sparsity structures in $B$ and $\Omega$ while allowing for uncertainty quantification using the full posterior. To this end, we propose to use the horseshoe prior that achieves efficient shrinkage in both sparse regression and inverse covariance estimation. We also develop an MCMC algorithm for sampling, without user-chosen tuning parameters.

\section{Proposed Model and Estimation Algorithm}
\label{sec:model}	

We define $\beta$ to be the vectorized coefficient matrix, or $\beta = vec(B') = [B_{11}, ..., B_{1q}, ..., B_{p1}, ..., B_{pq}]'$. To achieve shrinkage of the regression coefficients, we put horseshoe prior on $\beta$. That is,
	\begin{align*}
		\beta_j &\sim \textrm{N}(0,\lambda_j^2 \tau^2);\;  j=1,...,pq,  \\
		\lambda_{j} &\sim C^+(0,1),\; \tau \sim C^+(0,1),
	\end{align*}
 where $C^+(0,1)$ denotes the standard half-Cauchy distribution with density $p(x) \propto (1+x^2)^{-1}; \ x>0$. The normal scale mixture on $\beta$ with half-Cauchy hyperpriors on $\lambda_j$ and $\tau$ is known as the horseshoe prior \citep{carvalho2010horseshoe}, presumably due to the shape of the induced prior on the shrinkage factor. Similarly, to encourage sparsity in the off-diagonal elements of $\Omega$, we use the graphical horseshoe prior for Gaussian graphical models \citep{li2017graphical}, defined as,	
	\begin{align*}
		\omega_{kl:k>l} &\sim \textrm{N}(0,\eta_{kl}^2\zeta^2);\; k,l=1,...,q, \\
		\eta_{kl} \sim C^+(0,1)&, \; \zeta \sim C^+(0,1), \; \omega_{kk} \propto \textrm{constant},
	\end{align*}
\noindent where $\Omega= \{\omega_{kl}\}$, and the prior mass is truncated to the space of $q\times q$ positive definite matrices $\mathcal{S}_q^{+}$. In this model, $\eta_{kl}$ and $\zeta$ induce shrinkage on the off-diagonal elements in $\Omega$.

MCMC samplers have been proposed for regressions using the horseshoe prior for the linear regression model with i.i.d. error terms \citep{makalic2016samplerHS, bhattacharya2016fast}. However, these samplers cannot be applied to the current problem due to the correlation structure in the error. To transform the data into a model where sampling is possible, we reshape the predictors and responses. Let $\tilde{y}=vec(\Omega^{1/2}Y')$, and $\tilde{X}=X \otimes \Omega^{1/2}$. Simple algebra shows that $\tilde y \sim \textrm{N}_{nq}(\tilde{X}\beta, I_{nq})$. In this way, the matrix variate normal regression problem is transformed into a multivariate normal regression problem, provided the current estimate of $\Omega$ is known. Next, given the current estimate of $B$, the graphical horseshoe sampler of \citet{li2017graphical} is leveraged to estimate $\Omega$.

A full Gibbs sampler for the above model is given in Algorithm~\ref{alg:HS-GHS}. Throughout, the shape--scale parameterization is used for all gamma and inverse gamma random variables. First, the coefficient matrix $B$ is sampled conditional on the precision matrix $\Omega$. We notice that the conditional posterior of $\beta$ is $\textrm{N}((\tilde{X}'\tilde{X}+\Lambda_*^{-1})^{-1}\tilde{X}'\tilde{Y},(\tilde{X}'\tilde{X}+\Lambda_*^{-1})^{-1})$, where $\Lambda_*=\mathrm{diag}(\lambda_j^2 \tau^2), j=1,...,pq$. However, sampling from this normal distribution is computationally expensive because it involves computing the inverse of the $pq \times pq$ dimensional matrix $(\tilde{X}'\tilde{X}+\Lambda_*^{-1})$, with complexity $O(p^3q^3)$. Luckily, sampling $\beta$ from this high-dimensional normal distribution can be solved by the fast sampling scheme proposed by \citet{bhattacharya2016fast}. The algorithm is exact with a complexity linear in $p$.

\begin{algorithm}[!t]
	\caption{The HS-GHS Sampler}
	\label{alg:HS-GHS}
	\begin{footnotesize}
		\begin{algorithmic}
			\Function{HS-GHS}{$X, Y, burnin,nmc$}
			\State Set $n,p$ and $q$ using $\mathrm{dim}(X) = n\times p$ and $\mathrm{dim}(Y)=n\times q$
			\State Initialize $\beta=\mathbf{0}_{p\times q}$ and $\Omega= {I}_q$
			
			\For {$i = 1$ to $burnin+nmc$}
			
			\State (1) Calculate $\tilde{y}=vec(\Omega^{1/2}Y')$, $\tilde{X}=X \otimes \Omega^{1/2}$\\
			\indent \indent \indent \texttt{\%\% Sample $\beta$ using the horseshoe}
			\State (2a) Sample $u \sim \textrm{N}_{pq}(0,\Lambda_*)$ and $\delta \sim \textrm{N}_{nq}(0,I_{nq})$ independently, where $\Lambda_*=\mathrm{diag}(\lambda_j^2\tau^2)$
			\State (2b) Take $v = \tilde{X}u+\delta$
			\State (2c) Solve $w$ from $(\tilde{X}\Lambda_*\tilde{X}'+I_{nq})w=\tilde{y}-v$
			\State (2d) Calculate $\beta=u+\Lambda_*\tilde{X}'w$
			\State (3) Sample $\lambda_j^2 \sim \textrm{InvGamma}(1,1/\nu_j+\beta_j^2/(2\tau^2))$, and $\nu_j \sim \textrm{InvGamma}(1,1+1/\lambda_j^2)$, for $j=1,...,pq$
			\State (4) Sample $\tau^2 \sim \textrm{InvGamma}((pq+1)/2,1/\xi+\Sigma_{j=1}^{pq}\beta_j^2/(2\lambda_j^2))$, and $\xi \sim \textrm{InvGamma}(1,1+1/\tau^2)$
			\State (5) Calculate $Y_{res}=Y-XB$ and $S = Y_{res}'Y_{res}$\\
			\indent \indent \indent \texttt{\%\% Sample $\Omega$ using the graphical horseshoe}
			
			\For{$k = 1$ to $q$}
			\State Partition matrices $\Omega$, $S$ to $(q-1)\times(q-1)$ upper diagonal blocks $\Omega_{(-k)(-k)}$, $S_{(-k)(-k)}$; $(q-1)\times 1$ 
			\State dimensional vectors $\omega_{(-k)k}$, $s_{(-k)k}$; and scalars $\omega_{kk}$, $s_{kk}$
			\State (6a) Sample $\gamma \sim \text{Gamma}(n/2+1, 2/{s}_{kk})$
			\State (6b) Sample $\upsilon \sim \text{N}(-C s_{(-k)k},C)$ where $C = ({s}_{kk}\Omega_{(-k)(-k)}^{-1} +\text{diag}(\eta_{(-k)k}\zeta^2)^{-1})^{-1}$ and $\eta_{(-k)k}$
			\State  \indent is a vector of length $(q-1)$ with entries $\eta_{lk}^2, l \ne k$
			\State (6c) Apply transformation: $\omega_{(-k)k} = \upsilon,\, {\omega}_{kk} = \gamma + \upsilon'\Omega_{(-k)(-k)}^{-1}\upsilon$		
			\State (7) Sample $\eta_{(-k)k} \sim \text{InvGamma}(1,1/ \rho_{(-k)k}+\omega_{(-k)k}^2/2\zeta^2)$, 
			\State \indent and $\rho_{(-k)k} \sim \text{InvGamma}(1,\, 1+1/ \eta_{(-k)k})$ 
			\EndFor
			
			\State (8) Sample $\zeta^2 \sim \text{InvGamma}(({q\choose 2}+1)/2, 1/\phi+\sum_{k,l:k<l}{\omega_{kl}^2/2\eta_{kl}^2})$, and $\phi \sim \text{InvGamma}(1, 1+1/\zeta^2)$
			
			\State Save samples if $i > burnin$
			
			\EndFor
			
			\State Return MCMC samples of $\beta$ and $\Omega$
			
			\EndFunction
		\end{algorithmic}
	\end{footnotesize}
\end{algorithm}

To sample the precision matrix $\Omega$ conditional on $B$, define the residual $Y_{res}=Y-XB$, and let $S=Y_{res}'Y_{res}$. Since $Y -XB \sim \textrm{MN}(0, \ I_{n}, \ \Omega^{-1})$, the problem of estimating $\Omega$ given $B$ is exactly the zero-mean multivariate Gaussian inverse covariance estimation that the graphical horseshoe \citep{li2017graphical} solves. A detailed derivation of Algorithm~\ref{alg:HS-GHS} is given in Appendix~\ref{app:MCMC} and a MATLAB implementation, along with a simulation example, is freely available from github at \url{https://github.com/liyf1988/HS\_GHS}.

Complexity analysis of the proposed algorithm is as follows. Once $\Omega^{1/2}$ is calculated in $O(q^3)$ time, calculating $\tilde{y}$ costs $O(nq^2)$, and calculating $\tilde{X}$ costs $O(npq^2)$. The most time consuming step is still sampling $\beta$, which is $O(n^2pq^3)$ with the fast sampling method. Nevertheless, when $n \ll p$, using the fast sampling method is considerably less computationally intensive than sampling from the multivariate normal distribution directly, which has complexity $O(p^3q^3)$. Since the complexity of the graphical horseshoe is $O(q^3)$, each iteration in our Gibbs sampler takes $O(n^2pq^3)$ time.

Although the Gibbs sampler is computation-intensive, especially compared to penalized likelihood methods, it has several advantages. First, the Gibbs sampler is automatic, and does not require cross validation or empirical Bayes methods for choosing hyperparameters. Penalized optimization methods for simultaneous estimation of mean and inverse covariance usually need two tuning parameters \citep{cai2012covariate,rothman2010sparse,yin2011sparse}. Second, MCMC approximation of the the posterior distribution enables variable selection using posterior credible intervals. By varying the length of credible intervals, it is also possible to assess trade-offs between false positives and false negatives in variable selection. Finally, to our knowledge this is the first fully Bayesian solution in an SUR framework with a complexity linear in $p$. Along with these computational advantages, we now proceed to demonstrate the proposed method possesses attractive theoretical properties as well.

\section{Kullback--Leibler Risk Bounds}
\label{sec:KL}
Since a Bayesian method is meant to approximate an entire distribution, we provide results on Kullback--Leibler divergence between the true density (assuming there exists one) and the Bayes marginal density. Adopt the slightly non-Bayesian view that $n$ conditionally independent observations $Y_1,\ldots, Y_n$ are available from an underlying true parametric model with parameter $\theta_0$ and let $p^n$ denote the \emph{true joint density}, i.e., $p^n = \prod_{i=1}^{n} p(y_i; \theta_0)$. Similarly, let the marginal $m^n$ in a Bayesian model with prior $\nu(d\theta)$ on the parameter be defined as $m^n = \int \prod_{i=1}^{n} q(y_i | \theta) \nu (d\theta)$, where $q$ is the \emph{sampling density}. If the prior on $\theta$ is such that the measure of any set according to the true density and the sampling density are not too different, then it is natural to expect $p^n$ and $m^n$ to merge in information as more samples are available. The following result by \cite{barron1988exponential} formalizes this statement. Let $D_n(\theta)=\frac{1}{n} D(p^n||q^n(\cdot|\theta))$, where $D(\pi_1 | \pi_2 )=\int  \mathrm{log}(\pi_1/\pi_2) d\pi_1$, denotes the Kullback--Leibler divergence (KLD) of density $\pi_1$ with respect to $\pi_2$ and $q^n(\cdot|\theta)=\prod_{i=1}^{n} q(y_i | \theta)$. The set $A_\epsilon=\{\theta: D_n(\theta) <\epsilon\}$ can be thought of as a K--L information neighborhood of size $\epsilon$, centered at $\theta_0$. Then we have an upper bound on the KLD of $p^n$ from $m^n$, in terms of the prior measure of the set $D_n$.
\begin{lemma}
	\label{lm:barron}
	\citep{barron1988exponential}. Suppose the prior measure of the Kullback--Leibler information neighborhood is not exponentially small, i.e. for every $\epsilon$, $r>0$ there is an $N$ such that for all $n>N$ one has $\nu(A_\epsilon) \geq e^{-nr}$. Then:
	\begin{equation*}
		\frac{1}{n}D(p^n||m^n) \leq \epsilon-\frac{1}{n}\mathrm{log} \, \nu (A_\epsilon).
	\end{equation*}
\end{lemma}
The left hand side is the average Kullback--Leibler divergence between the true joint density of the samples $Y_1,...,Y_n$ and the marginal density. The right hand side involves logarithm of the prior measure of a Kullback--Leibler information neighborhood centered at $\theta_0$. A larger prior measure in this neighborhood of the ``truth'' gives a smaller upper bound for the average Kullback--Leibler divergence on the left, ensuring $p^n$ and $m^n$ are close in information. The following theorem shows that the HS-GHS prior, which has unbounded density at zero, achieves a smaller upper bound on the KLD when the true parameter is sparse (i.e., contains many zero elements), since it puts higher prior mass in an $\epsilon$ neighborhood of zero compared to any other prior with a bounded density at zero.
	
\begin{theorem}
	\label{thm:bound}
		Let $\theta_0= (B_0, \Omega_0)$ and assume $n$ conditionally independent observations $Y_1, \ldots, Y_n$ from the true model $Y_i \stackrel{ind} \sim \textnormal{N}(X_iB_0,\Omega_0^{-1})$, where $B_0\in \mathbb{R}^{p\times q}$ and $\Omega_0 \in \mathcal{S}_q^{+}$ are the true regression coefficients and inverse covariance, respectively and $X_i$ are observed covariates. Let $\beta_{j0}$, $\omega_{kl0}$ and $\sigma_{kl0}$ denote the $j$th and $kl$th element of $vec(B_0)$, $\Omega_0$ and $\Sigma_0 = \Omega_0^{-1}$, respectively. Suppose that $\sum_{k,l}\omega_{kl0} \propto q$, $\sum_{k,l}\sigma_{kl0} \propto q$, and $\sum_{i=1}^n (X_{i1}+\ldots+X_{ip})^2 \propto np^2$. Suppose that an Euclidean cube in the neighborhood of $\Omega_0$ with $(\omega_{kl0}-2/Mn^{1/2}q, \omega_{kl0}+2/Mn^{1/2}q)$ on each dimension lies in the cone of positive definite matrices $\mathcal{S}^{+}_q$, where $M=\sum_{k,l}\sigma_{kl0}/q$. Then, $ \frac{1}{n}D(p^n||m^n) \leq \frac{1}{n}-\frac{1}{n}\textnormal{log} \, \nu(A_{1/n})$ for all $n$, and:
	
	(1) For prior measure $\nu$ with density that is continuous, bounded above, and strictly positive in a neighborhood of zero, one obtains, $\mathrm{log} \, \nu(A_{1/n}) \propto K_1 pq \mathrm{log} (\frac{1}{n^{1/4}pq^{1/2}}) + K_2 q^2 \mathrm{log}(\frac{1}{n^{1/2}q}) $, where $K_1$ and $K_2$ are constants.
	
	(2) For prior measure $\nu$ under the HS-GHS prior, $\mathrm{log} \, \nu(A_{1/n}) > C_1 (pq-|s_B|) \mathrm{log} \{\frac{\mathrm{log}(n^{1/4}pq^{1/2})}{n^{1/4}pq^{1/2}} \} + C_2 |s_B| \mathrm{log} (\frac{1}{n^{1/4}pq^{1/2}}) + C_3 (q^2-|s_{\Omega}|) \mathrm{log}\{\frac{\mathrm{log}(n^{1/2}q)}{n^{1/2}q} \} + C_4 |s_{\Omega}| \mathrm{log}(\frac{1}{n^{1/2}q}) $, where $|s_B|$ is the number of nonzero elements in $B_0$, $|s_{\Omega}|$ is the number of nonzero elements in $\Omega_0$, and $C_1$, $C_2$, $C_3$, $C_4$ are constants.
	
\end{theorem}

Proof of Theorem~\ref{thm:bound} is in Appendix~\ref{app:bound}. Logarithm of the prior measure in the Kullback-Leibler divergence neighborhood, $\mathrm{log} \nu(A_{1/n})$, can be bounded by the summation of log measures in each of the $pq+q^2$ dimensions. Any Bayesian estimator with an element-wise prior satisfying conditions in Part (1) of Theorem~\ref{thm:bound} puts a prior measure proportional to $(n^{1/4}pq^{1/2})^{-1}$ in each of the $pq$ dimensions of the regression coefficients, and a measure proportional to $(n^{1/2}q)^{-1}$ in each of the $q^2$ dimensions of the inverse covariance, regardless of whether the corresponding true element is zero or non-zero. Theorem~\ref{thm:bound} implies that when $p$ and $q$ are fixed and $n \to \infty$, the average divergence $\frac{1}{n}D(p^n||m^n)$ under any Bayesian prior converges to zero. However, when $q$ is fixed and $p\mathrm{log}(n^{1/4}p)/n \to \infty$, the upper bound $n^{-1}\{1-\mathrm{log}\nu(A_{1/n}) \}$ diverges. Similarly, when $p$ is fixed and $q^2\mathrm{log}(n^{1/2}q)/n \to \infty$, the upper bound diverges. Some common Bayesian estimators, including the double exponential prior in Bayesian lasso, induce a prior density bounded above near the origin \citep{carvalho2010horseshoe}, satisfying conditions in Part (1). Being a mixture of double exponential priors, the spike-and-slab lasso prior also satisfies conditions in Part (1).

Although the upper bound diverges when $p$ and $q$ are large, it can be improved by putting higher prior mass near the origin when $B_0$ and $\Omega_0$ are sparse. One element where $\beta_{j0}=0$ contributes $\mathrm{log}(n^{1/4}pq^{1/2})/n$ to the upper bound under a bounded prior near the origin, and $\{\mathrm{log}(n^{1/4}pq^{1/2})-\mathrm{log}\,\mathrm{log}(n^{1/4}pq^{1/2})\}/n$ to the upper bound under the horseshoe prior. For each element where $\beta_{j0}=0$, the HS-GHS upper bound has an extra $-O\{(\mathrm{log}\,\mathrm{log}n^{1/4}pq^{1/2})/n\}$ term. Similarly, for each element where $\omega_{kl0}=0$, the HS-GHS upper bound has an extra $-O\{(\mathrm{log}\,\mathrm{log}n^{1/2}q)/n\}$ term. When most true coefficients and off-diagonal elements in the inverse covariance are zero, the horseshoe prior brings a non-trivial improvement on the upper bound. The theoretical findings of improved Kullback--Leibler divergence properties are extensively verified by simulations in Section~\ref{sec:sim}.

\section{Simulation Study}
\label{sec:sim}
In this section, we compare the performance of the HS-GHS prior to other multivariate normal regression methods that estimate both the regression coefficients and the precision matrix. We consider two cases, both with $p>n$. The first case has $p=200$ and $q=25$, and the second case has $p=120$ and $q=50$, and $n=100$ in both cases. We generate a sparse $p \times q$ coefficient matrix $B$ for each simulation setting, where $5\%$ of the elements in $B$ are nonzero. The nonzero elements in $B$ follow a uniform distribution in $(-2,-0.5) \bigcup (0.5,2)$. The precision matrix $\Omega$ is taken to be sparse with diagonal elements set to one and one of the following two patterns for off-diagonal elements:

\emph{1. AR1.} The precision matrix has an AR1 structure, with nonzero elements equal to 0.45.

\emph{2. Cliques.} The rows/columns are partitioned into disjoint groups and $\omega_{kl:k,l \in G, \ k \neq l}$ are set to $0.75$. When $q=25$, we consider eight groups and three members within each group. When $q=50$, the precision matrix contains $16$ groups and each group has three members.  It is important to note although these settings are used for the simulation examples, the proposed method allows arbitrary sparsity patterns in both $B$ and $\Omega$ and is in no way dependent on these specific settings.

We generate $n \times p$ design matrix $X$ with a toeplitz covariance structure where $Cov(X_i,X_j)=0.7^{|i-j|}$, and $n \times q$ error matrix $E \sim \mathrm{MN}(0,I_n, \Omega^{-1})$. The $n \times q$ response matrix is set to be $Y=XB+E$. For each simulation setting, $50$ data sets are generated, and $B$ and $\Omega$ are estimated by HS-GHS, MRCE \citep{rothman2010sparse}, CAPME \citep{cai2012covariate}, and the joint high-dimensional Bayesian variable and covariance selection (BM13) by \cite{bhadra2013joint}. The proposed HS-GHS estimator is implemented in MATLAB. The MATLAB code by \cite{bhadra2013joint} is used for BM13, and R packages `MRCE' and `capme' are used for MRCE and CAPME estimates. Mean squared estimation errors of the regression coefficients and the precision matrix; prediction mean squared error; average Kullback--Leibler divergence; and sensitivity (TP/(TP+FN)), specificity (TN/(TN+FP)), and precision (TP/(TP+FP)) in variable selection are reported. Here, TP, FP, TN and FN denote true positives, false positives, true negatives and false negatives, respectively. Variable selection for HS-GHS is performed using the middle $75\%$ posterior credible interval. Following \cite{bhadra2013joint}, variables with posterior probability of inclusion larger than $0.5$ are considered to be selected by BM13. In case the choices of these thresholds appear somewhat arbitrary, we also present receiver operating characteristic (ROC) curves for all methods to compare their overall variable selection performances as the decision threshold is varied between the two extremities, i.e., where all variables are selected and where none are selected.

Results are reported in Tables~\ref{tab:1} and \ref{tab:2}, along with CPU times for all methods. It is evident that the HS-GHS has the best overall statistical performance. Except for the mean squared error of $\Omega$ when $p=200$, the HS-GHS has the best estimation, prediction, information divergence and variable selection performances in our simulations. Although the HS-GHS does not have the highest sensitivity in recovering the support of $B$ or $\Omega$ in some cases, it has very high levels of specificity and precision. In other words, while the HS-GHS may miss some true signals, it finds far fewer false positives, so that a larger proportion of true positives exists in HS-GHS findings. This property of higher precision in identifying signals is an attractive feature in genomic applications.

In terms of the other methods, BM13 sometimes gives $\Omega$ estimate with the lowest mean squared error, but its estimate of $B$ has higher errors, and its sensitivity for recovering the support of $\Omega$ is low. MRCE estimation of $B$ is poor in higher dimensions, while CAPME has low mean squared errors in estimating both $B$ and $\Omega$. Both MRCE and CAPME are not stable in support recovery of $\Omega$. They either tend to select every element as a positive, giving high sensitivity and low specificity, or select every element as a negative, giving zero sensitivity and high specificity.

Figure~\ref{fig:ROC} shows the ROC curves for both $B$ and $\Omega$, when $p=120$ and $q=50$. True and false positive rates are generated by varying the width of posterior credible intervals from $0\%$ to $100\%$ in HS-GHS, and varying the posterior inclusion probability from $0\%$ to $100\%$ in BM13. In MRCE and CAPME, variables are selected by thresholding the estimated $B$ and $\Omega$. For each estimated $\beta_{j}$ and $\omega_{kl}$, the element is considered to be a positive if its absolute value is larger than a threshold, and the threshold is varied to generate a series of variable selection results. In all four plots, the HS-GHS curves closely follow the line where the true positive rate equals one, suggesting that the credible intervals for the true nonzero parameters do not include zero. These results are consistent with the theoretical findings that horseshoe credible intervals have optimal size \citep{van2017uncertainty}. CAPME has the second best performance in variable selection, except when it does not generate valid ROC plots. For example, in the cliques structured precision matrix estimated by CAPME, all off-diagonal elements are estimated to be zero, so CAPME cannot generate an ROC curve in this case. Moreover, neither MRCE nor BM13 produces satisfactory ROC curves. MCMC convergence diagnostics of the HS-GHS sampler are presented in Supplementary Section~\ref{sec:sup_mcmc} and further simulation results complementing the results in this section are in Supplementary Section~\ref{sec:sup_simulation}.

\section{Yeast eQTL Data Analysis}
\label{sec:yeast}

We illustrate the HS-GHS method using the yeast eQTL data analyzed by \cite{brem2005landscape}. The data set contains genome-wide profiling of expression levels and genotypes for $112$ yeast segregants from a cross between BY4716 and RM11-1a strains of \textit{Saccharomyces Cerevisiae}. This data set is available in the R package \texttt{trigger} (\url{https://www.bioconductor.org/packages/release/bioc/html/trigger.html}). The original data set contains expression values of $6216$ genes assayed on each array, and genotypes at $3244$ marker positions. Due to the small sample size, we only consider $54$ genes in the yeast mitogen-activated protein kinase (MAPK) signalling pathway in our analysis. This pathway was provided by the Kyoto Encyclopedia of Genes and Genomes database \citep{kanehisa2010kegg}, and was also analyzed by \cite{yin2011sparse} and \cite{cai2012covariate}.

Following the method described in \cite{curtis2013structured}, we divide the genome into $316$ groups based on linkage disequilibrium between the markers, and select the marker with the largest variation within each group. Then, we apply simple screening, and find $172$ markers that are marginally associated with at least one of the $54$ genes with a \textit{p}-value less than or equal to $0.01$. We use these $172$ markers as predictors and run a lasso regression on each of the $54$ genes. Residuals are used to assess the normality assumption. Based on qq-plots and normality tests, we drop five genes and two yeast segregants. Marginal qq-plots of residuals and other assessments of normality assumption are provided in Supplementary Section~\ref{sec:sup_normal}. The final data set we use in our analysis contains $49$ genes in the MAPK pathway and $172$ markers in $110$ yeast segregants.

We divide the $110$ yeast segregants into a training set containing $88$ segregants, and a testing set containing $22$ segregants. Coefficients of markers are estimated by HS-GHS, MRCE and CAPME using the training set, and the precision matrix of gene expressions are estimated as well. Prediction performance is measured over the testing set for each gene expression. Tuning parameters in MRCE and CAPME are selected by five-fold cross validation. Variable selection in HS-GHS are made by $75\%$ posterial credible interval. Prediction and estimation results are summarized in Tables~\ref{tab:pred} and \ref{tab:estim}, respectively. 

Out of $8428$ regression coefficients, CAPME estimates $182$ nonzero coefficients, MRCE estimates $11$ nonzero coefficients, and HS-GHS estimates $15$ nonzero coefficients. Prediction performance differs across these methods as well. For each gene expression, we use R-squared in the testing set, defined as $(1-$residual sum of squares/total sum of squares), to evaluate prediction. Many of the gene expressions cannot be predicted by any of the markers. Consequently, we only consider gene expressions that has R-squared larger than $0.1$ in any of these three models. Among $22$ such gene expressions, CAPME has the highest R-squared among the three methods in $4$ gene expressions, and HS-GHS has the highest R-squared in $18$ gene expressions. Average prediction R-squared values in these $22$ genes by CAPME, MRCE and HS-GHS are $0.1327$, $0.0063$, $0.2771$, respectively.

We also examine the $15$ nonzero coefficients estimated by the HS-GHS. CAPME estimates eight of these $15$ coefficients to be nonzero, and CAPME estimates have smaller absolute magnitudes than the HS-GHS estimates. In HS-GHS estimates, the genes SWI4 and SSK2 are associated with three markers each, and FUS1 is associated with two markers. The remaining gene expressions are associated with zero or one marker. One marker on chromosome $3$, location $201166$ is associated with four gene expressions (SWI4, SHO1, BCK1, SSK2), and it has the largest effect sizes among HS-GHS and CAPME estimated coefficients. This location is also identified as an eQTL hot spot by \cite{zhu2008integrating}. In addition, a marker on chromosome $5$ and a marker on chromosome $14$ in HS-GHS nonzero estimates also correspond to two other eQTL hot spots given by \cite{zhu2008integrating}. All of these nonzero estimates correspond to expressions mapped far from the location of their gene of origin, and can be considered distant eQTLs. This highlights the need for a model to simultaneously accommodate expressions and markers on different genomic locations, rather than separate chromosome-specific eQTL analysis.

Out of the $1176$ possible pairs among $49$ genes, CAPME, MRCE, and HS-GHS estimate $702$, $6$, and $88$ pairs to have nonzero partial covariance, respectively. We only present the HS-GHS estimated graph in Figure~\ref{fig:graph}, while CAPME and MRCE results are in Supplementary Section~\ref{sec:sup_graph}. Vertex colors in the graph indicate functions of genes. A current understanding of how yeast genes in the MAPK pathway respond to environmental stress and cellular signals, along with the functions of these genes, is available \citep{mapk}. Figure~\ref{fig:graph} recovers some known structures in the MAPK pathway. For instance, STE4, STE18, GPA1, STE20, CDC42, DIG1, BEM1, FUS1, STE2, STE3 and MSG5 are involved in the yeast mating process, and they are linked in the HS-GHS estimate. SLT2, SWI3, RHO1, RLM1 and MLP1 involved in the cell wall remodeling process, and YPD1, CTT1, GLO1 and SSK1 involved in the osmolyte synthesis process are also linked. It is also known that the high-osmolarity glycerol (HOG) and cell wall integrity (CWI) signalling pathways interact in yeast \citep{rodriguez2010high}, and some genes in the HOG pathway are indeed connected to genes in the CWI pathway in the HS-GHS estimate.

\section{Conclusions}
\label{sec:conclusion}

The horseshoe estimator has been shown to possess many attractive theoretical properties in sparse high-dimensional regressions. In this paper, we propose the HS-GHS estimator that generates sparse estimates of regression coefficients and inverse covariance simultaneously in multivariate Gaussian regressions. We implement the estimator using a full Gibbs sampler. Simulations in high-dimensional problems confirm that the HS-GHS outperforms several popular alternative methods in terms of estimation of both regression coefficients and inverse covariance, and in terms of prediction. The proposed method allows arbitrary sparsity patterns $B$ and $\Omega$ (as opposed to, say, methods based on decomposable graphs) and the number of unknown parameters inferred is $pq + q(q+1)/2$, which is indeed much larger than $n$ in all our examples. This fact needs to be accounted for before a na\"ive comparison of the scalability of SUR approaches with marginal correlation based methods for separate eQTL analysis. With $q=1$, the latter approaches may scale to larger values of $p$, but cannot utilize the error correlation structure as the SUR models do, consequently resulting in less statistically efficient estimates. HS-GHS also recovers the support of the regression coefficients and inverse covariance with higher precision compared to other SUR model based approaches, such as MRCE, CAPME and BM13. The proposed method is applied to yeast eQTL data for finding loci that explain genetic variation within the MAPK pathway, and identification of the gene network within this pathway.

The proposed method leverages and combines the beneficial properties of the horseshoe and graphical horseshoe priors, resulting in improved statistical performance. Computationally, the proposed sampler is the first in an SUR setting with a complexity linear in $p$, although the complexity is cubic in $q$. A major advantage of the proposed method is samples are available from the full posterior distribution, thereby allowing straightforward uncertainty quantification. If draws from the full posterior are not desired, it is possible faster algorithms can be developed to obtain point estimates. Prominent among these possibilities is an iterated conditional modes (ICM) algorithm \citep{besag1986statistical} that can be used to obtain the maximum pseudo posterior estimate. At each iteration, ICM maximizes the full conditional posteriors of all variables and converges to a deterministic solution. Since the full conditionals in the HS-GHS model are either normal, gamma or inverse gamma, the conditional modes are unique, and ICM should be easy to implement. It is also possible to include domain knowledge, such as pathway information, in the priors by coupling the local shrinkage parameters. This article focused on the horseshoe prior, which is a member of a broader class of global-local priors, sharing a sharp peak at zero and heavy tails. Performance of other priors belonging to this family, such as the horseshoe+ \citep{bhadra2017}, should also be explored. 

\section*{Supplementary Material}
The Supplementary Material contains MCMC convergence diagnostics and additional simulation results, referenced in Section~\ref{sec:sim} and additional results on eQTL data analysis, referenced in Section~\ref{sec:yeast}.

\section*{Acknowledgements}
Bhadra is supported by Grant No. DMS-1613063 by the US National Science Foundation.

\appendix

\section{Derivation of Algorithm \ref{alg:HS-GHS}}
\label{app:MCMC}
\begin{itemize}
\item \emph{Step 2:}  Since $\tilde{y} \sim \mathrm{N}_{nq}(\tilde{X}\beta,I_{nq})$ and the prior on $\beta$ is horseshoe, the full conditional posterior of $\beta$ is $\mathrm{N}((\tilde{X}'\tilde{X}+\Lambda_*^{-1})^{-1}\tilde{X}'\tilde{y},(\tilde{X}'\tilde{X}+\Lambda_*^{-1})^{-1})$, where $\Lambda_*=\mathrm{diag}(\lambda_j^2 \tau^2),j=1,...,pq$. Sampling of $\beta$ is exactly the problem solved by \cite{bhattacharya2016fast}. Realizing that $\beta$ has length $pq$, $\tilde{y}$ has length $nq$, and substituting $\tilde{X}$, $\tilde{y}$, $\Lambda_*$ and $\beta$ into Steps $1$ to $4$ in \textit{Algorithm 1} in \cite{bhattacharya2016fast}, yield Steps $(2a)$--$(2d)$.
	
\item \emph{Steps 3--4:} These steps concern sampling of the shrinkage parameters $\lambda_j$ for  $j=1,...,pq$, and $\tau$. Both have half Cauchy priors, which can be written as a mixture of two inverse gamma random variables. Specifically, if $x^2\mid a \sim \mathrm{InvGamma}(1/2,1/a)$ and $a \sim \mathrm{InvGamma}(1/2,1)$, then \cite{makalic2016samplerHS} demonstrated that marginally $x \sim \mathrm{C^+}(0,1)$. Since an inverse gamma prior is conjugate to itself and to the variance parameter in a normal model, the full conditional posteriors of $\lambda_j^2, \tau^2$ and the corresponding auxiliary variables $\nu_j$ and $\xi$ are all inverse gamma random variables. This completes Steps $3$ and $4$ in our Algorithm~\ref{alg:HS-GHS}.

\item \emph{Steps 6--8:} Given $B$, if one defines $Y_{res}=Y-XB$, then sampling of $\Omega$ is the problem of sampling the precision matrix in a zero-mean multivariate normal model. Thus, Steps (6a)--(8) in Algorithm~\ref{alg:HS-GHS} follows the sampling scheme of the graphical horseshoe model for sample size $n$, number of features $q$, and scatter matrix $S=Y_{res}'Y_{res}$. Details for these steps can be found in Algorithm 1 of \cite{li2017graphical}. 
\end{itemize}

\section{Proof of Theorem~\ref{thm:bound}}
\label{app:bound}

Let $A_{\epsilon} = \{\{B,\Omega\}: \frac{1}{n} D_n(p_{B_0,\Omega_0}||p_{B,\Omega}) \leq \epsilon \}$. We claim that $A_{\epsilon} \subset \mathbb{R}^{p\times q} \times \mathbb{R}^{q \times q}$ is bounded by an Euclidean cube of $pq+q^2$ dimensions with $(\beta_{j0}-k_1 \epsilon^{1/4}/pq^{1/2},\beta_{j0}+k_1 \epsilon^{1/4}/pq^{1/2})$, and $(\omega_{kl0}-k_2 \epsilon^{1/2}/q,\omega_{kl0}+k_2 \epsilon^{1/2}/q)$ on each dimension. The proof is as following.

Let $B = B_0 + (\epsilon^{1/4}/pq^{1/2}) \mathbb{1}_{p\times q}$, $\Omega = \Omega_0 + (\epsilon^{1/2}/q) \mathbb{1}_{q\times q}$, where $\mathbb{1}_{m \times n}$ denotes a $m \times n$ matrix with all elements equal to $1$. Then,
\begin{align*}
	D_n(p_{B_0,\Omega_0}||p_{B,\Omega}) =& \frac{n}{2} \{ \mathrm{log}|\Omega^{-1}\Omega_0| + tr(\Omega \Omega_0^{-1}) -q \} + \frac{1}{2}vec(XB-XB_0)'(\Omega \otimes I_n)vec(XB-XB_0) \\
	:=& \RN{1} + \RN{2}.
\end{align*}
By the proof of Theorem 3.2 in \cite{li2017graphical}, $\RN{1} \propto n\epsilon$ when $\epsilon \to 0$. We will show that $\RN{2} \propto n\epsilon$ as well. The expression for $\RN{2}$ is simplified as,
\begin{align*}
\RN{2} =& \frac{1}{2}vec(XB-XB_0)'(\Omega \otimes I_n)vec(XB-XB_0) \\
=& \frac{1}{2} \frac{\epsilon^{1/4}}{pq^{1/2}}vec(X\mathbb{1}_{p \times q})'\left\{\left(\Omega_0+\frac{\epsilon^{1/2}}{q} \mathbb{1}_{q \times q}\right) \otimes I_n \right\} \frac{\epsilon^{1/4}}{pq^{1/2}}vec(X\mathbb{1}_{p \times q}) \\
=& \frac{1}{2} \frac{\epsilon^{1/2}}{p^2 q} vec(X\mathbb{1}_{p \times q})'\left\{\Omega_0 \otimes I_n + \left(\frac{\epsilon^{1/2}}{q}\mathbb{1}_{q \times q}\right) \otimes I_n\right\} vec(X\mathbb{1}_{p \times q}).
\end{align*}
Some algebra shows that $vec(X\mathbb{1}_{p \times q})'(\Omega_0 \otimes I_n)vec(X\mathbb{1}_{p \times q}) = \sum_{k,l} \omega_{kl0} \sum_i (X_{i1}+\ldots+X_{ip})^2$, and $vec(X\mathbb{1}_{p \times q})'(\mathbb{1}_{q \times q} \otimes I_n)vec(X\mathbb{1}_{p \times q}) = q^2 \sum_i (X_{i1}+\ldots+X_{ip})^2$. Therefore,
\begin{align*}
\RN{2} =& \frac{1}{2} \frac{\epsilon^{1/2}}{p^2 q} \left\{ \sum_{k,l} \omega_{kl0} \sum_i (X_{i1}+\ldots+X_{ip})^2 + \frac{\epsilon^{1/2}}{q} q^2 \sum_i (X_{i1}+\ldots+X_{ip})^2\right\} \\
=& \frac{1}{2} \frac{\epsilon^{1/2}}{p^2 q}(c_1 n p^2 q + c_2 \epsilon^{1/2} n p^2 q) \\
=& \frac{1}{2} (c_1 n \epsilon^{1/2} + c_2 n \epsilon).
\end{align*}
Combining $\RN{1}$ and $\RN{2}$, $\frac{1}{n} D_n(p_{B_0,\Omega_0}||p_{B,\Omega}) \propto \epsilon$ when $\epsilon \to 0$. We have proved that $A_{\epsilon}$ is bounded by cubes of $pq+q^2$ dimensions described above. Now that we find cubes that bound $A_{\epsilon}$, we will bound $\nu(A_{\epsilon})$ by the product of prior measures on each dimension of these cubes. For any prior measure with density $p(\beta_j)$ that is continuous, bounded above, and strictly positive on a neighborhood of the true $\beta_{j0}$, one has $\int_{\beta_{j0}-\epsilon^{1/4}/(pq^{1/2})}^{\beta_{j0}+\epsilon^{1/4}/(pq^{1/2})} p(\beta_j) d\beta_j \propto {\epsilon^{1/4}}/{(pq^{1/2}})$, since the density is bounded above. Similarly, $\int_{\omega_{kl0}-\epsilon^{1/2}/q}^{\omega_{kl0}+\epsilon^{1/2}/q} p(\omega_{kl}) d\omega_{kl} \propto {\epsilon^{1/2}}/{q}$, for any prior density $p(\omega_{kl})$ satisfying the conditions. Taking $\epsilon=1/n$, this gives $\mathrm{log} \nu(A_{1/n})$ in Part(1) of Theorem~\ref{thm:bound}. The horseshoe prior also satisfies conditions in (1) in dimensions where $\beta_{j0} \neq 0$ and $\omega_{kl0} \neq 0$, so the same measures hold for HS-GHS in nonzero dimensions.

Now we need prior measure of horseshoe prior on dimensions where $\beta_{j0}=0$ and $\omega_{kl0}=0$. Using bounds of horseshoe prior provided in \cite{carvalho2010horseshoe}, it has been established by \cite{li2017graphical} that $\int_0^{\epsilon^{1/2}/q} p(\omega_{kl}) d\omega_{kl} > c_3 {\mathrm{log}(\epsilon^{-1/2}q)}/{(\epsilon^{-1/2}q)}$. Similar calculations show that $\int_0^{\epsilon^{1/4}pq^{1/2}}p(\beta_j)d\beta_j > c_4 {\mathrm{log}(\epsilon^{-1/4}pq^{1/2})}/{(\epsilon^{-1/4}pq^{1/2})}$. Taking $\epsilon = 1/n$, this gives Part (2) of the theorem and completes the proof.

\bibliographystyle{biom}
\bibliography{HS_SUR}

\begin{thebibliography}{}

\bibitem[\protect\citeauthoryear{Banterle, Bottolo, Richardson, Ala-Korpela,
  J{\"a}rvelin, and Lewin}{Banterle et~al.}{2018}]{banterle2018sparse}
Banterle, M., Bottolo, L., Richardson, S., Ala-Korpela, M., J{\"a}rvelin,
  M.-R., and Lewin, A. (2018).
\newblock Sparse variable and covariance selection for high-dimensional
  seemingly unrelated {B}ayesian regression.
\newblock {\em bioRxiv, doi: 10.1101/467019} .

\bibitem[\protect\citeauthoryear{Barron}{Barron}{1988}]{barron1988exponential}
Barron, A.~R. (1988).
\newblock {\em The exponential convergence of posterior probabilities with
  implications for Bayes estimators of density functions}.
\newblock Technical report, Department of Statistics, University of Illinois,
  Champaign, IL.

\bibitem[\protect\citeauthoryear{Besag}{Besag}{1986}]{besag1986statistical}
Besag, J. (1986).
\newblock On the statistical analysis of dirty pictures.
\newblock {\em Journal of the Royal Statistical Society. Series B
  (Methodological)} {\bf 48,} 259--302.

\bibitem[\protect\citeauthoryear{Bhadra, Datta, Li, Polson, and Willard}{Bhadra
  et~al.}{2019}]{bhadra2016prediction}
Bhadra, A., Datta, J., Li, Y., Polson, N.~G., and Willard, B. (2019).
\newblock Prediction risk for the horseshoe regression.
\newblock {\em Journal of Machine Learning Research} {\bf 20,} 1--39.

\bibitem[\protect\citeauthoryear{Bhadra, Datta, Polson, and Willard}{Bhadra
  et~al.}{2017}]{bhadra2017}
Bhadra, A., Datta, J., Polson, N.~G., and Willard, B. (2017).
\newblock The horseshoe+ estimator of ultra-sparse signals.
\newblock {\em Bayesian Analysis} {\bf 12,} 1105--1131.

\bibitem[\protect\citeauthoryear{Bhadra and Mallick}{Bhadra and
  Mallick}{2013}]{bhadra2013joint}
Bhadra, A. and Mallick, B.~K. (2013).
\newblock Joint high-dimensional {B}ayesian variable and covariance selection
  with an application to e{QTL} analysis.
\newblock {\em Biometrics} {\bf 69,} 447--457.

\bibitem[\protect\citeauthoryear{Bhattacharya, Chakraborty, and
  Mallick}{Bhattacharya et~al.}{2016}]{bhattacharya2016fast}
Bhattacharya, A., Chakraborty, A., and Mallick, B.~K. (2016).
\newblock Fast sampling with {G}aussian scale mixture priors in
  high-dimensional regression.
\newblock {\em Biometrika} {\bf 103,} 985--991.

\bibitem[\protect\citeauthoryear{Brem and Kruglyak}{Brem and
  Kruglyak}{2005}]{brem2005landscape}
Brem, R.~B. and Kruglyak, L. (2005).
\newblock The landscape of genetic complexity across 5,700 gene expression
  traits in yeast.
\newblock {\em Proceedings of the National Academy of Sciences} {\bf 102,}
  1572--1577.

\bibitem[\protect\citeauthoryear{Cai, Li, Liu, and Xie}{Cai
  et~al.}{2012}]{cai2012covariate}
Cai, T.~T., Li, H., Liu, W., and Xie, J. (2012).
\newblock Covariate-adjusted precision matrix estimation with an application in
  genetical genomics.
\newblock {\em Biometrika} {\bf 100,} 139--156.

\bibitem[\protect\citeauthoryear{Candes and Tao}{Candes and
  Tao}{2007}]{candes2007dantzig}
Candes, E. and Tao, T. (2007).
\newblock The {D}antzig selector: Statistical estimation when $p$ is much
  larger than $n$.
\newblock {\em The Annals of Statistics} {\bf 35,} 2313--2351.

\bibitem[\protect\citeauthoryear{Carvalho, Polson, and Scott}{Carvalho
  et~al.}{2010}]{carvalho2010horseshoe}
Carvalho, C.~M., Polson, N.~G., and Scott, J.~G. (2010).
\newblock The horseshoe estimator for sparse signals.
\newblock {\em Biometrika} {\bf 97,} 465--480.

\bibitem[\protect\citeauthoryear{Conklin, Adriaens, Kelder, and
  Salomonis}{Conklin et~al.}{2018}]{mapk}
Conklin, B., Adriaens, M., Kelder, T., and Salomonis, N. (2018).
\newblock {MAPK} signaling pathway (saccharomyces cerevisiae).
\newblock \url{https://www.wikipathways.org/index.php/Pathway:WP510}.
\newblock [Online; accessed 12-December-2018].

\bibitem[\protect\citeauthoryear{Curtis, Kim, Woolford~Jr, Xu, and Xing}{Curtis
  et~al.}{2013}]{curtis2013structured}
Curtis, R.~E., Kim, S., Woolford~Jr, J.~L., Xu, W., and Xing, E.~P. (2013).
\newblock Structured association analysis leads to insight into saccharomyces
  cerevisiae gene regulation by finding multiple contributing e{QTL} hotspots
  associated with functional gene modules.
\newblock {\em BMC Genomics} {\bf 14,} 196.

\bibitem[\protect\citeauthoryear{Datta and Ghosh}{Datta and
  Ghosh}{2013}]{datta2013asymptotic}
Datta, J. and Ghosh, J.~K. (2013).
\newblock Asymptotic properties of {Bayes} risk for the horseshoe prior.
\newblock {\em Bayesian Analysis} {\bf 8,} 111--132.

\bibitem[\protect\citeauthoryear{Dawid}{Dawid}{1981}]{dawid1981some}
Dawid, A.~P. (1981).
\newblock Some matrix-variate distribution theory: notational considerations
  and a bayesian application.
\newblock {\em Biometrika} {\bf 68,} 265--274.

\bibitem[\protect\citeauthoryear{Dawid and Lauritzen}{Dawid and
  Lauritzen}{1993}]{dawid1993hyper}
Dawid, A.~P. and Lauritzen, S.~L. (1993).
\newblock Hyper markov laws in the statistical analysis of decomposable
  graphical models.
\newblock {\em The Annals of Statistics} {\bf 21,} 1272--1317.

\bibitem[\protect\citeauthoryear{Deshpande, Ro\v{c}kov\'a, and
  George}{Deshpande et~al.}{2019}]{deshpande2017simultaneous}
Deshpande, S.~K., Ro\v{c}kov\'a, V., and George, E.~I. (2019).
\newblock Simultaneous variable and covariance selection with the multivariate
  spike-and-slab lasso.
\newblock {\em Journal of Computational and Graphical Statistics} {\bf to
  appear,}.

\bibitem[\protect\citeauthoryear{Holmes, Denison, and Mallick}{Holmes
  et~al.}{2002}]{holmes2002accounting}
Holmes, C.~C., Denison, D.~T., and Mallick, B.~K. (2002).
\newblock Accounting for model uncertainty in seemingly unrelated regressions.
\newblock {\em Journal of Computational and Graphical Statistics} {\bf 11,}
  533--551.

\bibitem[\protect\citeauthoryear{Kanehisa, Goto, Furumichi, Tanabe, and
  Hirakawa}{Kanehisa et~al.}{2010}]{kanehisa2010kegg}
Kanehisa, M., Goto, S., Furumichi, M., Tanabe, M., and Hirakawa, M. (2010).
\newblock {KEGG} for representation and analysis of molecular networks
  involving diseases and drugs.
\newblock {\em Nucleic Acids Research} {\bf 38,} D355--D360.

\bibitem[\protect\citeauthoryear{Leclerc}{Leclerc}{2008}]{leclerc2008survival}
Leclerc, R.~D. (2008).
\newblock Survival of the sparsest: robust gene networks are parsimonious.
\newblock {\em Molecular Systems Biology} {\bf 4,} 213.

\bibitem[\protect\citeauthoryear{Li, Craig, and Bhadra}{Li
  et~al.}{2019}]{li2017graphical}
Li, Y., Craig, B.~A., and Bhadra, A. (2019).
\newblock The graphical horseshoe estimator for inverse covariance matrices.
\newblock {\em Journal of Computational and Graphical Statistics} {\bf to
  appear,}.

\bibitem[\protect\citeauthoryear{Makalic and Schmidt}{Makalic and
  Schmidt}{2016}]{makalic2016samplerHS}
Makalic, E. and Schmidt, D.~F. (2016).
\newblock A simple sampler for the horseshoe estimator.
\newblock {\em IEEE Signal Processing Letters} {\bf 23,} 179--182.

\bibitem[\protect\citeauthoryear{Meinshausen, Rocha, and Yu}{Meinshausen
  et~al.}{2007}]{meinshausen2007discussion}
Meinshausen, N., Rocha, G., and Yu, B. (2007).
\newblock Discussion: A tale of three cousins: Lasso, {L2B}oosting and
  {D}antzig.
\newblock {\em The Annals of Statistics} {\bf 35,} 2373--2384.

\bibitem[\protect\citeauthoryear{Rodr{\'\i}guez-Pe{\~n}a, Garc{\'\i}a, Nombela,
  and Arroyo}{Rodr{\'\i}guez-Pe{\~n}a et~al.}{2010}]{rodriguez2010high}
Rodr{\'\i}guez-Pe{\~n}a, J.~M., Garc{\'\i}a, R., Nombela, C., and Arroyo, J.
  (2010).
\newblock The high-osmolarity glycerol ({HOG}) and cell wall integrity ({CWI})
  signalling pathways interplay: a yeast dialogue between {MAPK} routes.
\newblock {\em Yeast} {\bf 27,} 495--502.

\bibitem[\protect\citeauthoryear{Rothman, Levina, and Zhu}{Rothman
  et~al.}{2010}]{rothman2010sparse}
Rothman, A.~J., Levina, E., and Zhu, J. (2010).
\newblock Sparse multivariate regression with covariance estimation.
\newblock {\em Journal of Computational and Graphical Statistics} {\bf 19,}
  947--962.

\bibitem[\protect\citeauthoryear{Schadt, Monks, Drake, Lusis, Che, Colinayo,
  Ruff, Milligan, Lamb, Cavet, et~al\mbox{.}}{Schadt
  et~al.}{2003}]{schadt2003genetics}
Schadt, E.~E., Monks, S.~A., Drake, T.~A., Lusis, A.~J., Che, N., Colinayo, V.,
  Ruff, T.~G., Milligan, S.~B., Lamb, J.~R., Cavet, G., et~al. (2003).
\newblock Genetics of gene expression surveyed in maize, mouse and man.
\newblock {\em Nature} {\bf 422,} 297--302.

\bibitem[\protect\citeauthoryear{Touloumis, Marioni, and Tavar\'{e}}{Touloumis
  et~al.}{2016}]{touloumis2016hdtd}
Touloumis, A., Marioni, J.~C., and Tavar\'{e}, S. (2016).
\newblock {HDTD}: analyzing multi-tissue gene expression data.
\newblock {\em Bioinformatics} {\bf 32,} 2193--2195.

\bibitem[\protect\citeauthoryear{van~der Pas, Kleijn, and van~der
  Vaart}{van~der Pas et~al.}{2014}]{van2014horseshoe}
van~der Pas, S., Kleijn, B., and van~der Vaart, A. (2014).
\newblock The horseshoe estimator: Posterior concentration around nearly black
  vectors.
\newblock {\em Electronic Journal of Statistics} {\bf 8,} 2585--2618.

\bibitem[\protect\citeauthoryear{van~der Pas, Szab{\'o}, and van~der
  Vaart}{van~der Pas et~al.}{2017}]{van2017uncertainty}
van~der Pas, S., Szab{\'o}, B., and van~der Vaart, A. (2017).
\newblock Uncertainty quantification for the horseshoe (with discussion).
\newblock {\em Bayesian Analysis} {\bf 12,} 1221--1274.

\bibitem[\protect\citeauthoryear{Yin and Li}{Yin and Li}{2011}]{yin2011sparse}
Yin, J. and Li, H. (2011).
\newblock A sparse conditional {G}aussian graphical model for analysis of
  genetical genomics data.
\newblock {\em The Annals of Applied Statistics} {\bf 5,} 2630--2650.

\bibitem[\protect\citeauthoryear{Zellner}{Zellner}{1962}]{zellner1962efficient}
Zellner, A. (1962).
\newblock An efficient method of estimating seemingly unrelated regressions and
  tests for aggregation bias.
\newblock {\em Journal of the American statistical Association} {\bf 57,}
  348--368.

\bibitem[\protect\citeauthoryear{Zheng and Liu}{Zheng and
  Liu}{2011}]{zheng2011experimental}
Zheng, S. and Liu, W. (2011).
\newblock An experimental comparison of gene selection by lasso and {D}antzig
  selector for cancer classification.
\newblock {\em Computers in Biology and Medicine} {\bf 41,} 1033--1040.

\bibitem[\protect\citeauthoryear{Zhu, Zhang, Smith, Drees, Brem, Kruglyak,
  Bumgarner, and Schadt}{Zhu et~al.}{2008}]{zhu2008integrating}
Zhu, J., Zhang, B., Smith, E.~N., Drees, B., Brem, R.~B., Kruglyak, L.,
  Bumgarner, R.~E., and Schadt, E.~E. (2008).
\newblock Integrating large-scale functional genomic data to dissect the
  complexity of yeast regulatory networks.
\newblock {\em Nature Genetics} {\bf 40,} 854--861.

\end{thebibliography}


\begin{table}[!t]
	\centering
	\vspace{5cm}
	\noindent\makebox[\textwidth]{%
	\begin{threeparttable}
	\caption{Mean squared error (sd) in estimation and prediction, average Kullback--Leibler divergence, and sensitivity, specificity and precision of variable selection performance, over 50 simulated data sets, $p=200$ and $q=25$. The regression coefficients and precision matrix are estimated by HS-GHS, joint high-dimensional Bayesian variable and covariance selection (BM13), MRCE and CAPME. The best performer in each column is shown in bold.}
	
	\label{tab:1}
	\begin{footnotesize}
			\begin{tabular}{l|ccc|c|ccc|ccc|c|}
			\toprule
				& \multicolumn{11}{|c|}{Simulation 1: $p=200$, $q=25$, $n=100$, Uniform coefficients, AR1 structure} \\
				& \multicolumn{3}{|c|}{MSE} &
				\multicolumn{1}{|c|}{Divergence} &
				\multicolumn{3}{|c|}{B support recovery} &
				\multicolumn{3}{|c|}{$\Omega$ support recovery} &
				CPU time \\
				\toprule
				Method & B & $\Omega$ & Prediction & avg KL &
				SEN & SPE & PRC & SEN & SPE & PRC & min. \\
				\toprule
				HS-GHS & \textbf{0.0033} & 0.0365 & \textbf{2.6352} & \textbf{10.2075} &
					.9380 & .9981 & \textbf{.9621} & \textbf{.9658} & .9973 & \textbf{.9700} & 
					788.75 \\
					& (0.0005) & (0.0123) & (0.1792) & (1.2853) &
					(.0155) & (.0006) & (.0122) & (.0383) & (.0039) & (.0418) & \\
				BM13 & 0.0560 & \textbf{0.0301} & 8.4230 & 14.8512 &
					- & - & - & .0200 & \textbf{.9986} & .5588 $^1$ &
					54.80 \\
					& (0.0006) & (0.0005) & (0.4276) & (0.3441) &
					- & - & - & (.0242) & (.0019) & (.4567) & \\
				MRCE & 0.0854 & 0.0476 & 19.4201 & 29.9000 &
					.0208 & \textbf{.9996} & .8074 & .9425 & .0907 & .0828 &
					0.28 \\
					& (0.0007) & (0.0006) & (0.8754) & (0.3824) &
					(.0083) & (.0004) & (.1751) & (.0733) & (.0724) & (.0028) & \\
				CAPME & 0.0156 & 0.0417 & 4.0337 & 12.1094 &
					\textbf{.9445} & .8187 & .2167 & 0 & 1 & - $^2$ &
					74.60 \\
					& (0.0014) & (0.0010) & (0.2749) & (0.4189) &
					(.0130) & (.0201) & (.0182) & (0) & (0) & - & \\	
			\toprule
				& \multicolumn{11}{|c|}{Simulation 2: $p=200$, $q=25$, $n=100$, Uniform coefficients, Cliques structure} \\
				& \multicolumn{3}{|c|}{MSE} &
				{Divergence} &
				\multicolumn{3}{|c|}{B support recovery} &
				\multicolumn{3}{|c|}{$\Omega$ support recovery} &
				CPU time \\
				\toprule
				Method & B & $\Omega$ & Prediction & avg KL &
				SEN & SPE & PRC & SEN & SPE & PRC & min. \\
				\toprule
				HS-GHS & \textbf{0.0058} & \textbf{0.0371} & \textbf{3.5388} & \textbf{9.0762} &
					.8696 & .9985 & \textbf{.9693} & \textbf{.9700} & .9972 & \textbf{.9687} &
					788.31 \\
					& (0.0010) & (0.0253) & (0.1791) & (1.3446) &
					(.0204) & (.0008) & (.0159) & (.0430) & (.0030) & (.0331) & \\
				BM13 & 0.0570 & 0.0595 & 9.2452 & 14.3267 &
					- & - & - & .0204 & \textbf{.9993} & .7500 $^3$ &
					54.79 \\
					& (0.0006) & (0.0006) & (0.4789) & (0.4324) &
					- & - & - & (.0242) & (.0014) & (.3808) & \\
				MRCE & 0.0861 & 0.0756 & 20.1694 & 27.3668 &
					.0116 & \textbf{.9999} & .9370 & .9507 & .0788 & .0825 &
					0.16 \\
					& (0.0005) & (0.0006) & (0.9440) & (0.2892) &
					(.0057) & (.0001) & (.1121) & (.0581) & (.0596) & (.0041) & \\
				CAPME & 0.0188 & 0.0718 & 5.0170 & 11.2598 &
					\textbf{.9266} & .8270 & .2218 & 0 & 1 & - $^4$ &
					73.67 \\
					& (0.0016) & (0.0007) & (0.2930) & (0.3797) &
					(.0155) & (.0215) & (.0198) & (0) & (0) & - & \\
			\bottomrule
		\end{tabular}
		\begin{tablenotes}
			\item 1. 16 NaNs in 50 replicates. 3. 23 NaNs in 50 replicates. 2,4. 50 NaNs. All mean and sd. calculated on non-NaN values.
		\end{tablenotes}
	\end{footnotesize}
	\end{threeparttable}}
\end{table}

\newpage

\begin{table}[!t]
	\centering
	\noindent\makebox[\textwidth]{%
		\begin{threeparttable}
			\caption{Mean squared error (sd) in estimation and prediction, average Kullback--Leibler divergence, and sensitivity, specificity and precision of variable selection performance, over 50 simulated data sets, $p=120$ and $q=50$. The regression coefficients and precision matrix are estimated by HS-GHS, joint high-dimensional Bayesian variable and covariance selection (BM13), MRCE and CAPME. The best performer in each column is shown in bold.}
			
			\label{tab:2}
			\begin{footnotesize}
				\begin{tabular}{l|ccc|c|ccc|ccc|c|}
					\toprule
					& \multicolumn{11}{|c|}{Simulation 3: $p=120$, $q=50$, $n=100$, Uniform coefficients, AR1 structure} \\
					& \multicolumn{3}{|c|}{MSE} &
					{Divergence} &
					\multicolumn{3}{|c|}{B support recovery} &
					\multicolumn{3}{|c|}{$\Omega$ support recovery} &
					CPU time \\
					\toprule
					Method & B & $\Omega$ & Prediction & avg KL &
					SEN & SPE & PRC & SEN & SPE & PRC & min. \\
					\toprule
					HS-GHS & \textbf{0.0022} & \textbf{0.0041} & \textbf{2.4495} & \textbf{8.0596} &
					\textbf{.9709} & \textbf{.9984} & \textbf{.9696} & \textbf{.9873} & \textbf{.9995} & \textbf{.9875} &
					2.57e+03 \\
					& (0.0002) & (0.0009) & (0.1055) & (0.6494) &
					(.0087) & (.0007) & (.0120) & (.0136) & (.0007) & (.0156) & \\
					BM13 & 0.0493 & 0.0132 & 5.1923 & 25.1810 &
					- & - & - & .2804 & .9976 & .8295 &
					217.24 \\
					& (0.0006) & (0.0006) & (0.2091) & (0.7590) &
					- & - & - & (.0603) & (.0015) & (.1058) & \\
					MRCE & 0.0689 & 0.0150 & 10.5162 & 40.3985 &
					.2774 & .9897 & .5895 & .9755 & .1218 & .0442 &
					10.34 \\
					& (0.0022) & (0.0004) & (0.5920) & (0.8349) &
					(.0281) & (.0023) & (.0431) & (.0189) & (.0116) & (.0009) & \\
					CAPME & 0.0151 & 0.0105 & 3.2662 & 14.6163 &
					.9462 & .8887 & .3122 & .9514 & .9795 & .6705 $^1$ &
					80.69 \\
					& (0.0015) & (0.0013) & (0.1501) & (0.9668) &
					(.0131) & (.0184) & (.0280) & (.1390) & (.0093) & (.0782) & \\
					\toprule
					& \multicolumn{11}{|c|}{Simulation 4: $p=120$, $q=50$, $n=100$, Uniform coefficients, Cliques structure} \\
					& \multicolumn{3}{|c|}{MSE} &
					{Divergence} &
					\multicolumn{3}{|c|}{B support recovery} &
					\multicolumn{3}{|c|}{$\Omega$ support recovery} &
					CPU time \\
					\toprule
					Method & B & $\Omega$ & Prediction & avg KL &
					SEN & SPE & PRC & SEN & SPE & PRC & min. \\
					\toprule
					HS-GHS & \textbf{0.0032} & \textbf{0.0052} & \textbf{3.0221} & \textbf{7.8564} &
					.9409 & \textbf{.9986} & \textbf{.9717} & \textbf{.9992} & .9990 & \textbf{.9776} &
					2.57e+03 \\
					& (0.0004) & (0.0028) & (0.0983) & (0.8065) &
					(.0131) & (.0006) & (.0121) & (.0059) & (.0013) & (.0284) & \\
					BM13 & 0.0506 & 0.0290 & 5.8167 & 24.0404 &
					- & - & - & .0904 & \textbf{.9993} & .8414 &
					216.83 \\
					& (0.0007) & (0.0005) & (0.2225) & (0.6104) &
					- & - & - & (.0359) & (.0007) & (.1497) & \\
					MRCE & 0.0774 & 0.0298 & 12.0456 & 41.3306 &
					.1527 & .9971 & .7398 & .9679 & .0940 & .0419 &
					8.06 \\
					& (0.0014) & (0.0010) & (0.6366) & (0.7870) &
					(.0192) & (.0009) & (.0625) & (.0684) & (.0780) & (.0020) & \\
					CAPME & 0.0161 & 0.0331 & 3.8324 &  16.9539 &
					\textbf{.9537} & .8373 & .2384 & 0 & 1 & - $^2$ &
					81.99 \\
					& (0.0013) & (0.0004) & (0.1421) & (0.4293) &
					(.0122) & (.0234) & (.0251) & (0) & (0) & - & \\				
					\bottomrule
				\end{tabular}
				\begin{tablenotes}
					\item 1. 1 NaN in 50 replicates. 2. 50 NaNs. Mean and sd. calculated on non-NaN values.
				\end{tablenotes}
			\end{footnotesize}
	\end{threeparttable}}
\end{table}

\newpage

\begin{table}[!t]
	\centering
	\vspace{7cm}
	\caption{Percentage of model explained variation in prediction of gene expressions. Model coefficients are estimated in training set $(n=88)$ and prediction performance is evaluated in testing set $(n=22)$.}
	\label{tab:pred}
	\begin{footnotesize}
	\noindent\makebox[\textwidth]{%
		\begin{tabular}{l|ccc||l|ccc}
			\toprule
			Gene & CAPME & MRCE & HS-GHS & Gene & CAPME & MRCE & HS-GHS \\
			\toprule
			FUS3 & \textbf{15.46} & 0.00 & 2.12 & TEC1 & 23.08 & 0.00 & \textbf{26.27} \\
			FUS1 & \textbf{31.78} & 0.00 & 17.60 & SSK22 & 21.24 & 0.00 & \textbf{59.57} \\
			STE2 & 43.78 & 0.00 & \textbf{79.76} & MF(ALPHA)2 & 23.64 & 0.00 & \textbf{48.27} \\
			GPA1 & \textbf{19.50} & 0.00 & 1.38 & FAR1 & \textbf{30.66} & 0.00 & 1.47 \\
 			STE3 & 36.19 & 0.00 & \textbf{76.45} & MF(ALPHA)1 & 39.37 & 0.00 & \textbf{80.93} \\
			BEM1 & 0.00 & 0.00 & \textbf{16.68} & STE5 & 0.00 & 4.90 & \textbf{19.60} \\
			KSS1 & 2.80 & 0.00 & \textbf{21.76} & SLN1 & 4.38 & 0.00 & \textbf{10.41} \\
			STE18 & 0.00 & 0.00 & \textbf{24.88} & MLP1 & 0.00 & 0.00 & \textbf{10.19} \\
			HOG1 & 0.00 & 0.00 & \textbf{19.28} & FKS1 & 0.00 & 0.00 & \textbf{32.09} \\
			MCM1 & 0.00 & 0.00 & \textbf{29.96} & WSC3 & 0.00 & 0.00 & \textbf{10.20} \\
			SLG1 & 0.00 & 8.98 & \textbf{10.27} & RHO1 & 0.00 & 0.00 & \textbf{10.57} \\
			\bottomrule
		\end{tabular}}
	\end{footnotesize}
		\vspace{10cm}
\end{table}

\newpage
\begin{table}[!t]
	\centering
	\noindent\makebox[\textwidth]{%
	\begin{threeparttable}
		\caption{Nonzero coefficients in HS-GHS estimate, along with names and locations of the genes, locations of the markers, and CAPME estimated coefficients.}
	\label{tab:estim}
	\begin{footnotesize}
			\begin{tabular}{ccccccc}
				\toprule
				Gene & Chromosome & Within-chr. & Marker chr. & Within-chr. & HS-GHS & CAPME \\
				& & position & & marker position & coefficients & coefficients \\
				\toprule
				FUS3 & 2 & 192454-193515 & 2 & 424330 & 0.32 & 0.06 \\
				BEM1 & 2 & 620867-622522 & 8 & 71742 & -0.35 & 0.00 \\
				FUS1 & 3 & 71803-73341 & 4 & 17718 & 0.13 & 0.00 \\
				FUS1 & 3 & 71803-73341 & 4 & 527445 & -0.42 & -0.13 \\
				SWI4 & 5 & 382591-385872 & 13 & 361370 & -0.88 & 0.00 \\
				SWI4 & 5 & 382591-385872 & 5 & 458085 & -0.69 & 0.00 \\
				SWI4 & 5 & 382591-385872 & 3 & 201166 & 3.65 & 2.00 \\
				SHO1 & 5 & 397948-399051 & 3 & 201166 & -1.89 & -0.91 \\
				BCK1 & 10 & 247250-251686 & 3 & 201166 & -4.11 & -2.66 \\
				MID2 & 12 & 790676-791806 & 13 & 314816 & 0.29 & 0.06 \\
				STE11 & 12 & 849865-852018 & 5 & 109310 & 0.13 & 0.00 \\
				MFA2 & 14 & 352416-352532 & 14 & 449639 & 0.13 & 0.00$^{1}$ \\
				SSK2 & 14 & 680696-685435 & 5 & 395442 & 0.98 & 0.00 \\
				SSK2 & 14 & 680696-685435 & 13 & 403766 & 0.68 & 0.08 \\
				SSK2 & 14 & 680696-685435 & 3 & 201166 & -3.60 & -2.05 \\
				\bottomrule				
			\end{tabular}
		\begin{tablenotes}
			{\item 1. MRCE estimate for this coefficient is 0.05 and MRCE estimates for all other coefficients in this table are 0. Thus, MRCE results are not separately presented.}
		\end{tablenotes}
		\end{footnotesize}
\end{threeparttable}}
\end{table}
\begin{figure*}[t]
	\centering
	\begin{subfigure}[t]{0.45\textwidth}
		\centering
		\adjincludegraphics[width=\textwidth,trim={{0.12\width} {0.05\height} {0.16\width} {0\height}},clip]{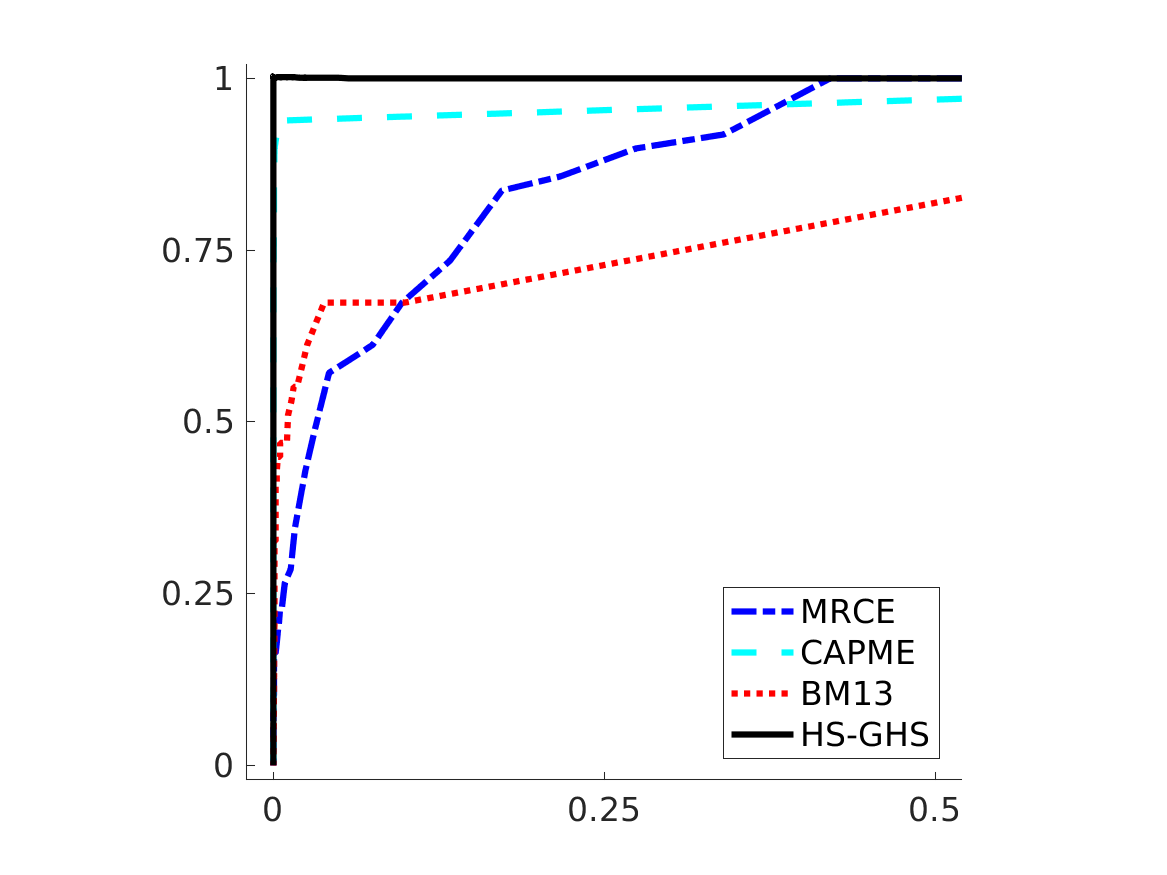}
		\caption[ROC]%
		{{\footnotesize Support recovery of $\Omega$, AR1 structure}}
		\label{fig:ROC_omega_ar1}
	\end{subfigure}
	\
	\begin{subfigure}[t]{0.45\textwidth}
		\centering
		\adjincludegraphics[width=\textwidth,trim={{0.12\width} {0.05\height} {0.16\width} {0\height}},clip]{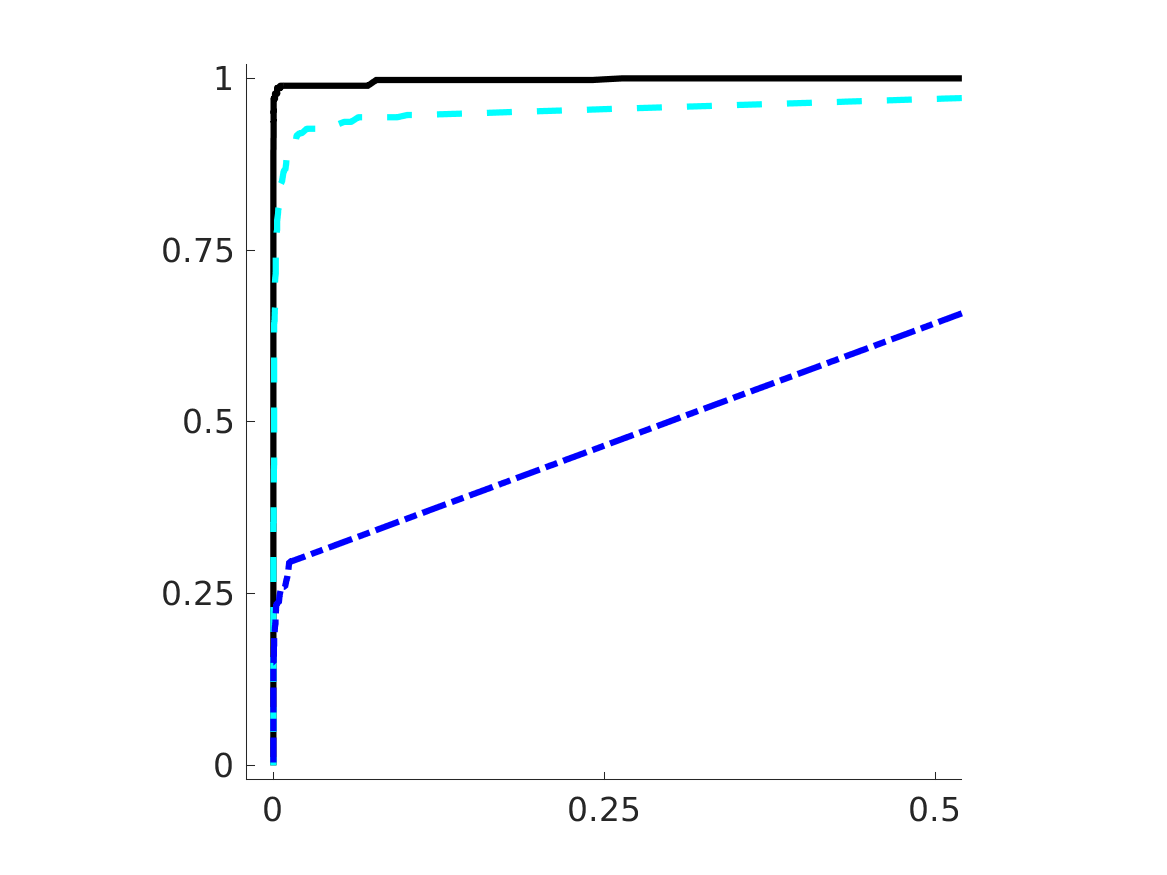}
		\caption[]%
		{{\footnotesize Support recovery of $B$ with AR1 structure in $\Omega$}}
		\label{fig:ROC_B_ar1}
	\end{subfigure}
	\
	\begin{subfigure}[t]{0.45\textwidth}
		\centering
		\adjincludegraphics[width=\textwidth,trim={{0.12\width} {0.05\height} {0.16\width} {0\height}},clip]{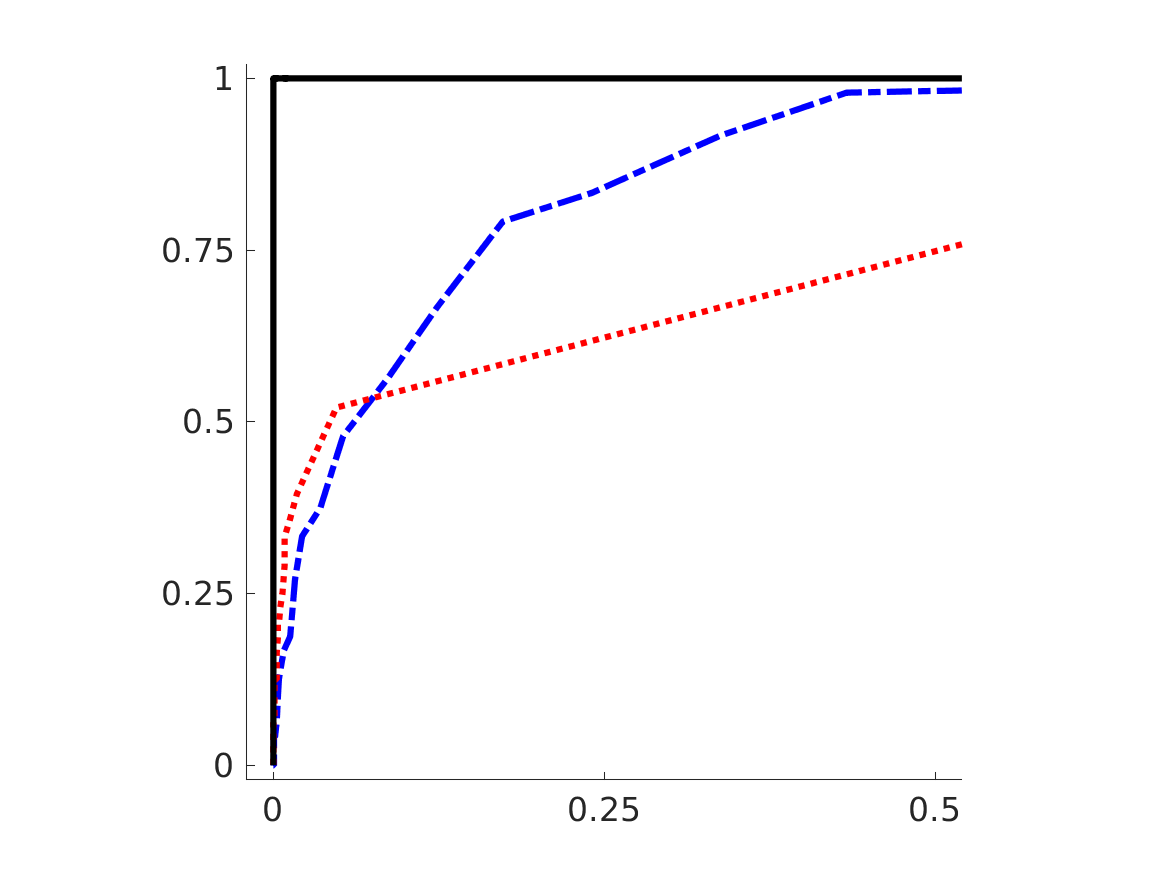}
		\caption[]%
		{{\footnotesize Support recovery of $\Omega$, Cliques structure}}
		\label{fig:ROC_omega_clique}
	\end{subfigure}
	\
	\begin{subfigure}[t]{0.45\textwidth}
		\centering
		\adjincludegraphics[width=\textwidth,trim={{0.12\width} {0.05\height} {0.16\width} {0\height}},clip]{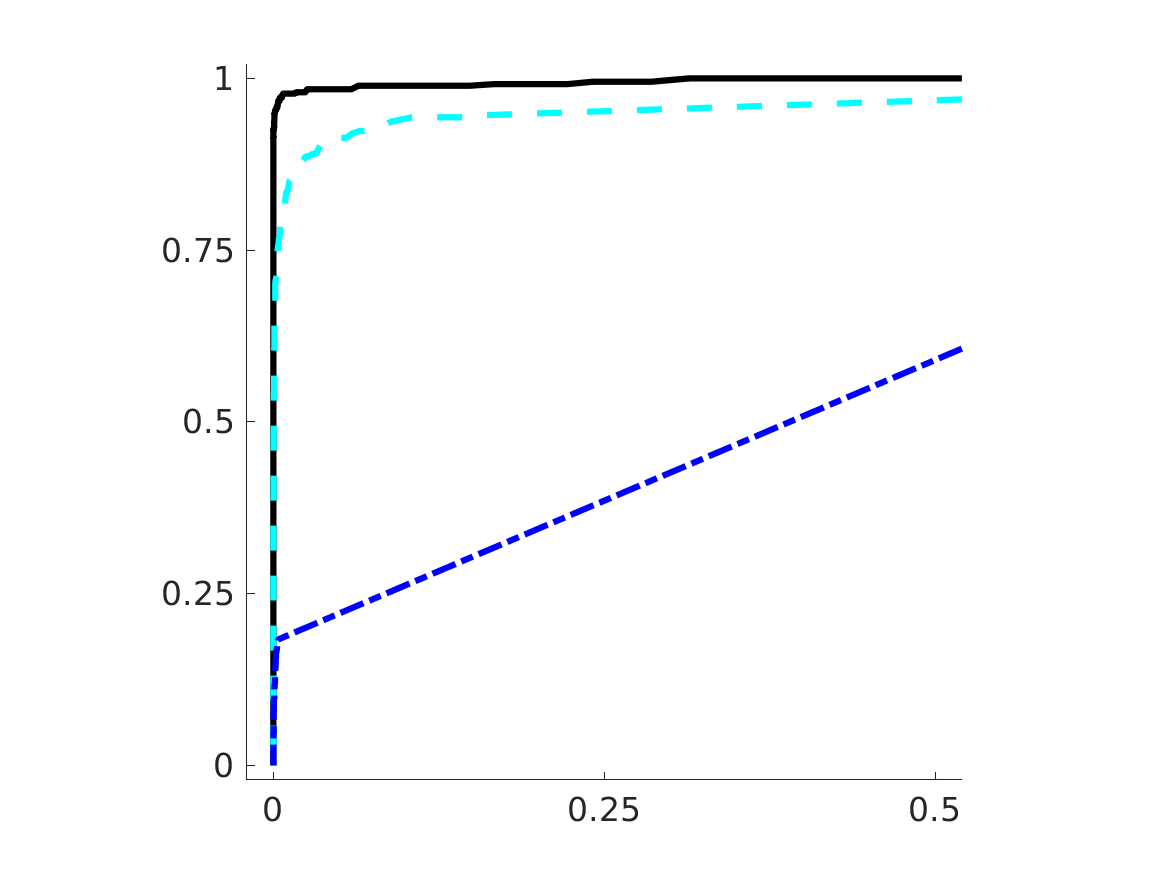}
		\caption[]%
		{{\footnotesize Support recovery of $B$ with Cliques structure in $\Omega$}}
		\label{fig:ROC_B_clique}
	\end{subfigure}
	\caption{Receiver operating characteristic (ROC) curves of estimates by HS-GHS, joint high-dimensional Bayesian variable and covariance selection (BM13), MRCE and CAPME for $p=120$ and $q=50$. The true positive rates are shown on the y-axis, and the false positive rates are shown on the x-axis.}
	\label{fig:ROC}
\end{figure*}

\newpage
\ \ 
\begin{figure}[!t]
	\centering
		\adjincludegraphics[width=0.85\textwidth,trim={{0.1\width} {0.1\height} {0.1\width} {0.1\height}},clip]{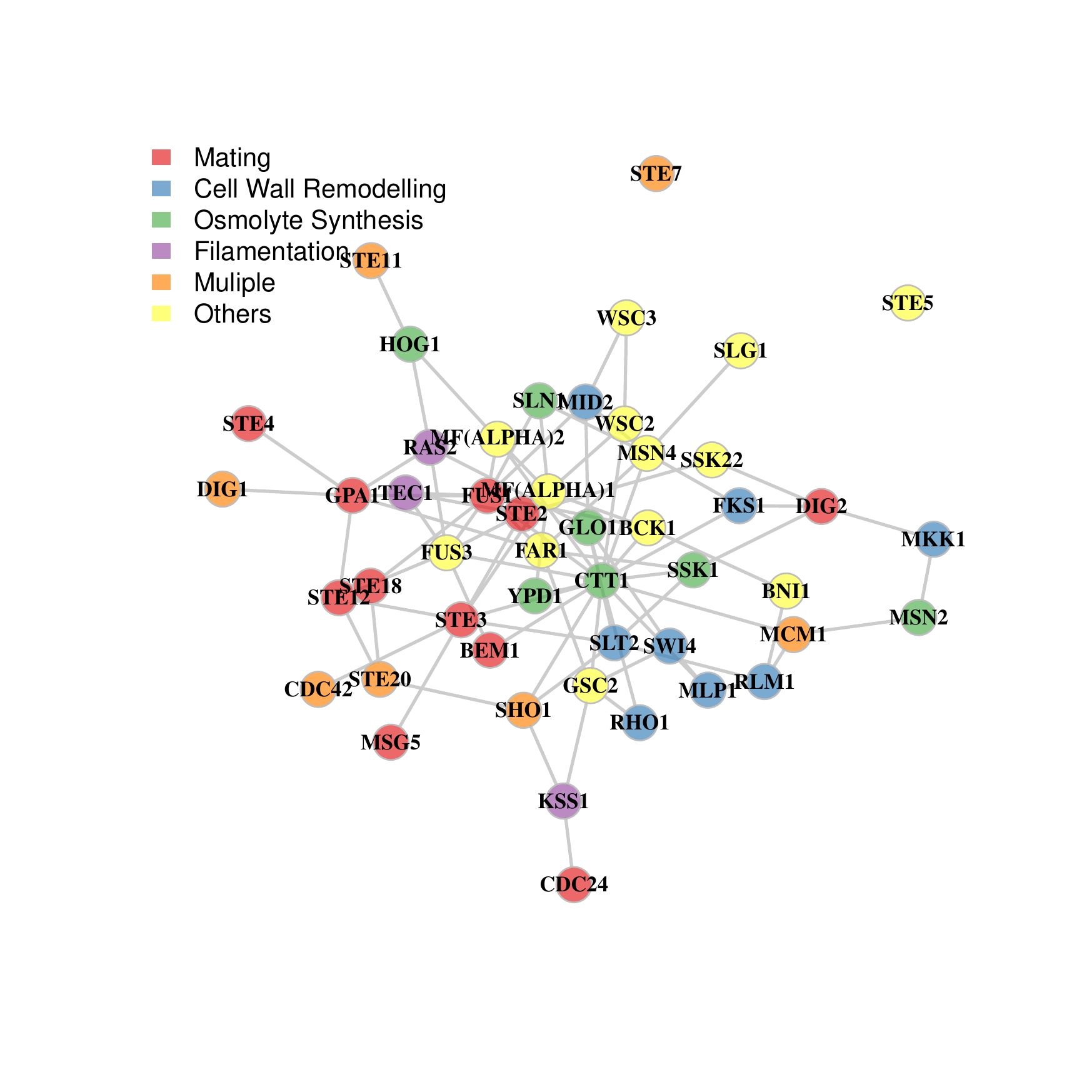}
	\caption{The inferred graph for gene expressions in the MAPK pathway by the HS-GHS estimate. Vertex colors indicate functions of genes.}
	\label{fig:graph}
\end{figure}

\newpage
\clearpage\pagebreak\newpage
\thispagestyle{empty}
\begin{center}
{\LARGE{\bf Supplementary Material to\\ {\it Joint Mean--Covariance Estimation via the Horseshoe with an Application in Genomic Data Analysis}}}
\end{center}
\vskip 2cm
\baselineskip=15pt
\begin{center}
\vspace{-1cm}
Yunfan Li\\
Department of Statistics, Purdue University\\
\hskip 5mm \\
Jyotishka Datta\\
Department of Mathematical Sciences, University of Arkansas\\
\hskip 5mm \\
Bruce A. Craig\\
Department of Statistics, Purdue University\\
\hskip 5mm \\
Anindya Bhadra\\
Department of Statistics, Purdue University\\
\end{center}

\setcounter{equation}{0}
\setcounter{page}{0}
\setcounter{table}{0}
\setcounter{section}{0}
\setcounter{subsection}{0}
\setcounter{figure}{0}
\renewcommand{\theequation}{S.\arabic{equation}}
\renewcommand{\thesection}{S.\arabic{section}}
\renewcommand{\thesubsection}{S.\arabic{subsection}}
\renewcommand{\thepage}{S.\arabic{page}}
\renewcommand{\thetable}{S.\arabic{table}}
\renewcommand{\thefigure}{S.\arabic{figure}}

\newpage

\subsection{MCMC Convergence Diagnostics}
\label{sec:sup_mcmc}

\begin{figure*}[h]
	\centering
	\begin{subfigure}[t]{0.45\textwidth}
		\centering
		\adjincludegraphics[width=\textwidth,trim={{0.075\width} {0\height} {0.05\width} {0\height}},clip]{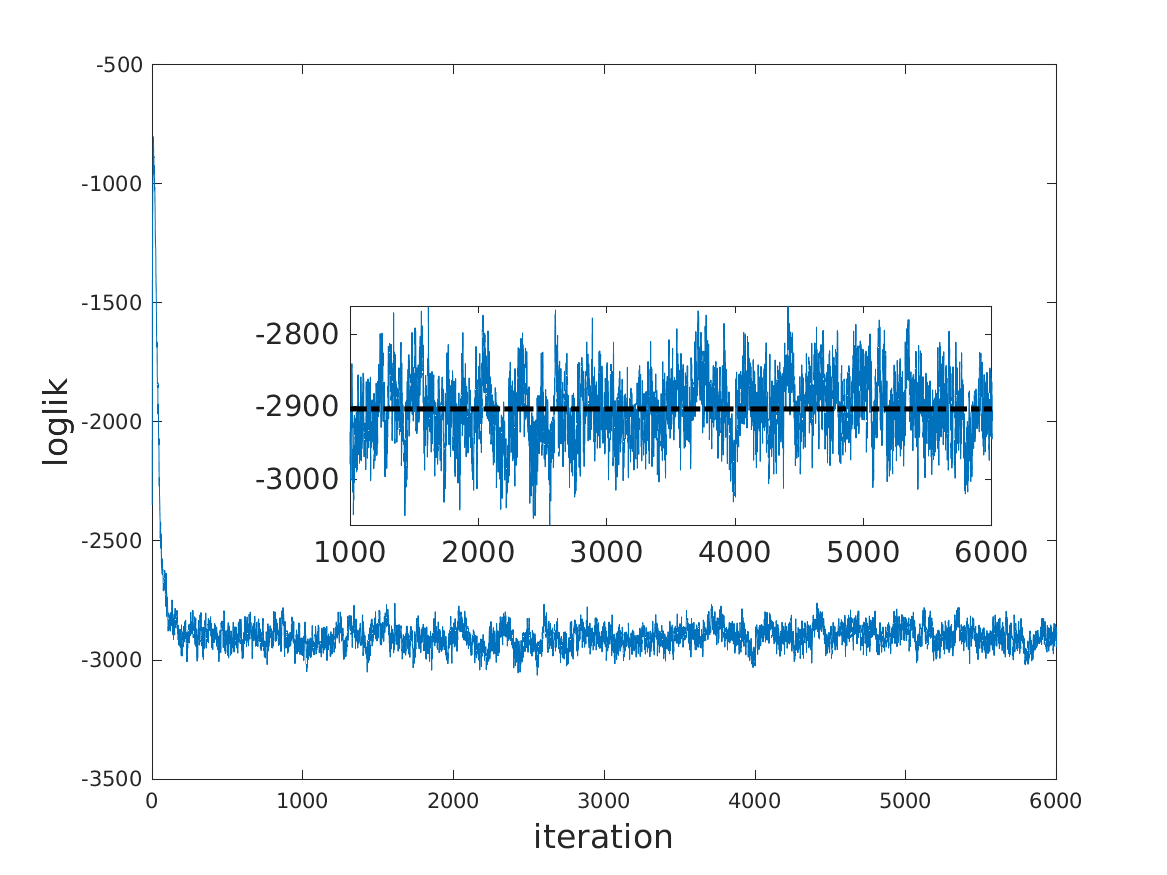}
		\caption[ROC]%
		{{\footnotesize Log likelihood vs iteration, $p=120$, $q=50$}}
		\label{fig:loglik_p120q50}
	\end{subfigure}
	\qquad
	\begin{subfigure}[t]{0.45\textwidth}
		\centering
		\adjincludegraphics[width=\textwidth,trim={{0.075\width} {0\height} {0.05\width} {0\height}},clip]{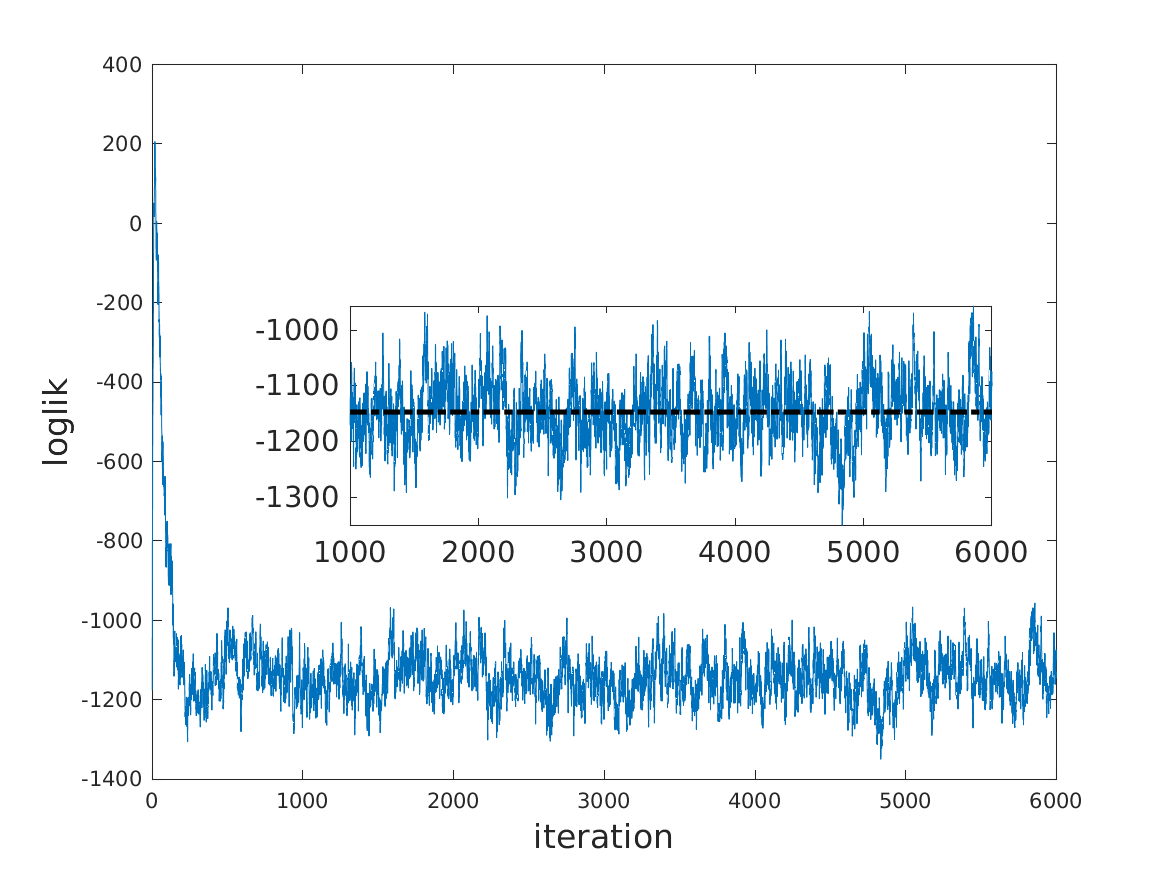}
		\caption[ROC]%
		{{\footnotesize Log likelihood vs iteration, $p=200$, $q=25$}}
		\label{fig:loglik_p200q25}
	\end{subfigure}
	\caption{Log likelihood at each iteration using Algorithm~\ref{alg:HS-GHS} for HS-GHS, under (a) AR1 structured inverse covariance matrix, $p=120$, $q=50$, and (b) AR1 structured inverse covariance matrix, $p=200$, $q=25$. Horizontal lines show log likelihood averaged over iterations $1000$ to $6000$. The first data set in the corresponding simulations are used. \label{fig:loglik}}	
\end{figure*}
Figure~\ref{fig:loglik} shows the trace plots of the log likelihood over 6,000 MCMC iterations and the inside panel in each plot shows the trace plot after discarding the first 1,000 draws as burn-in samples. The plots indicate quick mixing. Formal MCMC diagnostics, such as the Gelman--Rubin test, could be performed using the MCMC output, if desired.

\subsection{Additional Simulation Results}
\label{sec:sup_simulation}

We provide additional simulation results, complementing those in Section~\ref{sec:sim}. Tables~\ref{stab:1} and \ref{stab:2} provide results when $p=100$, $q=25$. Tables~\ref{stab:3} and \ref{stab:4} supplement Tables~\ref{tab:1} and \ref{tab:2} with more simulation settings. In the star structured inverse covariance matrix, $\omega_{1k}=0.25, \, k=2,...,q$, all diagonal elements equal to $1$, and the rest of the elements all equal to $0$. In the case of large coefficients, all nonzero coefficients are equal to $5$. Other structures of the inverse covariance matrix and uniformly distributed coefficients are described in Section~\ref{sec:sim}. One-fifth of the coefficients are nonzero when $p=100$ and $q=25$, and $5\%$ of the coefficients are nonzero in the other dimensions.

\begin{table}[H]
	\centering
	\noindent\makebox[\textwidth]{%
	\begin{threeparttable}
	\caption{Mean squared error (sd) in estimation and prediction, average Kullback--Leibler divergence, and sensitivity, specificity and precision of variable selection performance, over 50 simulated data sets. The regression coefficients and precision matrix are estimated by HS-GHS, joint high-dimensional Bayesian variable and covariance selection (BM13), MRCE and CAPME. The best performer in each column is shown in bold.}
	
	\label{stab:1}
	\begin{footnotesize}
			\begin{tabular}{l|ccc|c|ccc|ccc|c|}
				\toprule
				& \multicolumn{11}{|c|}{Simulation 1: $p=100$, $q=25$, $n=100$, Uniform coefficients, AR1 structure} \\
				& \multicolumn{3}{|c|}{MSE} &
				{Divergence} &
				\multicolumn{3}{|c|}{B support recovery} &
				\multicolumn{3}{|c|}{$\Omega$ support recovery} &
				CPU time \\
				\toprule
				Method & B & $\Omega$ & Prediction & avg KL &
				SEN & SPE & PRC & SEN & SPE & PRC & min. \\
				\toprule
				HS-GHS & \textbf{0.0166} & 0.0047 & \textbf{2.6235} & \textbf{4.6663} &
				.9674 & \textbf{.9695} & \textbf{.8885} & .9942 & \textbf{.9959} & \textbf{.9565} &
				96.53 \\
				& (0.0017) & (0.0015) & (0.1740) & (0.4478) &
				(.0136) & (.0060) & (.0196) & (.0146) & (.0034) & (.0356) & \\
				BM13 & 0.1396 & 0.0313 & 4.9680 & 12.7152 &
				- & - & - & .5533 & .9903 & .8363 & 
				6.17 \\
				& (0.0035) & (0.0012) & (0.3073) & (0.3328) &
				- & - & - & (.0758) & (.0051) & (.0760) & \\
				MRCE & 0.0230 & \textbf{0.0034} & 2.7459 & 5.0754 &
				\textbf{.9952} & .6373 & .4076 & \textbf{.9992} & .8249 & .3399 &
				24.27 \\
				& (0.0022) & (0.0011) & (0.1851) & (0.4761) &
				(.0045) & (.0267) & (.0178) & (.0059) & (.0416) & (.0543) & \\
				CAPME & 0.0460 & 0.0253 & 3.2043 & 8.4143 &
				.9761 & .5775 & .3704 & .5075 & .9801 & .7184$^1$ &
				40.06 \\
				& (0.0061) & (0.0100) & (0.2287) & (1.3079) &
				(.0141) & (.0747) & (.0382) & (.4931) & (.0294) & (.1354) & \\
				\toprule
				& \multicolumn{11}{|c|}{Simulation 2: $p=100$, $q=25$, Uniform coefficients, Star structure} \\
				& \multicolumn{3}{|c|}{MSE} &
				{Divergence} &
				\multicolumn{3}{|c|}{B support recovery} &
				\multicolumn{3}{|c|}{$\Omega$ support recovery} &
				CPU time \\
				\toprule
				Method & B & $\Omega$ & Prediction & avg KL &
				SEN & SPE & PRC & SEN & SPE & PRC & min. \\
				\toprule
				HS-GHS & \textbf{0.0138} & \textbf{0.0058} & \textbf{1.5459} & \textbf{4.6722} &
				.9789 & \textbf{.9630} & \textbf{.8693} & .5089 & .9955 & \textbf{.8882} &
				96.62 \\
				& (0.0014) & (0.0018) & (0.0856) & (0.4392) &
				(.0100) & (.0069) & (.0211) & (.1540) & (.0051) & (.1153) & \\
				BM13 & 0.1362 & 0.0188 & 3.8594 & 12.2304 &
				- & - & - & .0289 & .9943 & .2307$^2$ &
				4.91 \\
				& (0.0034) & (0.0004) & (0.2270) & (0.1689) &
				- & - & - & (.0359) & (.0037) & (.2708) & \\
				MRCE & 0.0193 & 0.0109 & 1.6357 & 6.2894 &
				.9938 & .6270 & .4004 & \textbf{.9167} & .8575 & .3356$^3$ &
				21.17 \\
				& (0.0021) & (0.0033) & (0.0863) & (0.6595) &
				(.0051) & (.0252) & (.0159) & (.1761) & (.0731) & (.1439) & \\
				CAPME & 0.0255 & 0.0143 & 1.8071 & 5.6583 &
				\textbf{.9954} & .5099 & .3377 & 0 & \textbf{1} & -$^4$ &
				40.33 \\
				& (0.0026) & (0.0012) & (0.1016) & (0.2677) &
				(.0043) & (.0379) & (.0174) & (0) & (0) & - & \\
				\bottomrule
		\end{tabular}
		\begin{tablenotes}
			\item 1.23 NaNs in 50 replicates. 2. 5 NaNs in 50 replicates. 3. 1 NaN in 50 replicates. 4. 50 NaNs. All mean and sd. calculated on non-NaN values.
		\end{tablenotes}
	\end{footnotesize}
	\end{threeparttable}}
\end{table}

\begin{table}[H]
	\centering
	\noindent\makebox[\textwidth]{%
		\begin{threeparttable}
			\caption{Mean squared error (sd) in estimation and prediction, average Kullback--Leibler divergence, and sensitivity, specificity and precision of variable selection performance, over 50 simulated data sets. The regression coefficients and precision matrix are estimated by HS-GHS, joint high-dimensional Bayesian variable and covariance selection (BM13), MRCE and CAPME. The best performer in each column is shown in bold.}
			
			\label{stab:2}
			\begin{footnotesize}
				\begin{tabular}{l|ccc|c|ccc|ccc|c|}
					\toprule
					& \multicolumn{11}{|c|}{Simulation 3: $p=100$, $q=25$, Uniform coefficients, Cliques structure} \\
					& \multicolumn{3}{|c|}{MSE} &
					{Divergence} &
					\multicolumn{3}{|c|}{B support recovery} &
					\multicolumn{3}{|c|}{$\Omega$ support recovery} &
					CPU time \\
					\toprule
					Method & B & $\Omega$ & Prediction & avg KL &
					SEN & SPE & PRC & SEN & SPE & PRC & min. \\
					\toprule
					HS-GHS & \textbf{0.0261} & \textbf{0.0044} & \textbf{3.3854} & \textbf{4.5777} &
					.9148 & \textbf{.9701} & \textbf{.8846} & \textbf{1} & .9952 & .9499 &
					96.53 \\
					& (0.0026) & (0.0018) & (0.1745) & (0.4076) &
					(.0188) & (.0057) & (.0194) & (0) & (.0044) & (.0437) & \\
					BM13 & 0.1417 & 0.0601 & 5.5897 & 11.4533 &
					- & - & - & .4567 & .9988 & \textbf{.9674} &
					4.80 \\
					& (0.0038) & (0.0021) & (0.3192) & (0.3326) &
					- & - & - & (.1163) & (.0019) & (.0611) & \\
					MRCE & 0.0363 & 0.0147 & 3.5770 & 6.4222 &
					\textbf{.9763} & .6443 & .4079 & \textbf{1} & .6924 & .2293 &
					24.55 \\
					& (.0036) & (.0057) & (.1968) & (0.7096) &
					(.0110) & (.0300) & (.0198) & (0) & (.0815) & (.0457) & \\
					CAPME & 0.0534 & 0.0668 & 3.9208 & 8.8355 &
					.9697 & .5535 & .3538 & 0 & \textbf{1} & -$^1$ &
					40.78 \\
					& (0.0053) & (0.0009) & (0.2257) & (0.2505) &
					(.0119) & (.0530) & (.0260) & (0) & (0) & - & \\
					\toprule
					& \multicolumn{11}{|c|}{Simulation 4: $p=100$, $q=25$, Coefficients=5, AR1 structure} \\
					& \multicolumn{3}{|c|}{MSE} &
					{Divergence} &
					\multicolumn{3}{|c|}{B support recovery} &
					\multicolumn{3}{|c|}{$\Omega$ support recovery} &
					CPU time \\
					\toprule
					Method & B & $\Omega$ & Prediction & avg KL &
					SEN & SPE & PRC & SEN & SPE & PRC & min. \\
					\toprule
					HS-GHS & \textbf{0.0125} & \textbf{0.0057} & \textbf{2.5508} & \textbf{4.5945} &
					\textbf{1} & \textbf{.9669} & \textbf{.8836} & \textbf{.9950} & .9954 & \textbf{.9514} &
					97.66 \\
					& (0.0013) & (0.0018) & (0.1705) & (0.5065) &
					(0) & (.0074) & (.0233) & (.0137) & (.0039) & (.0401) & \\
					BM13 & 1.5774 & 0.0521 & 35.2494 & 34.3200 &
					- & - & - & .2133 & .9659 & .3533 &
					4.85 \\
					& (0.0325) & ($<$0.0001) & (2.5042) & (0.2394) &
					- & - & - & (.0732) & (.0086) & (.1029) & \\
					MRCE & 0.0550 & 0.0113 & 3.3325 & 11.9130 &
					\textbf{1} & .1830 & .2346 & .9900 & .8510 & .3965 &
					27.45 \\
					& (0.0094) & (0.0066) & (0.2599) & (2.1085) &
					(0) & (.0396) & (.0090) & (.0332) & (.0648) & (.1161) & \\
					CAPME & 0.0638 & 0.0377 & 4.5765 & 14.4007 &
					\textbf{1} & .5498 & .3588 & 0 & \textbf{1} & -$^2$ &
					38.93 \\
					& (0.0086) & (0.0013) & (0.5602) & (1.4711) &
					(0) & (.0491) & (.0256) & (0) & (0) & - & \\
					
					\bottomrule
				\end{tabular}
				\begin{tablenotes}
					\item 1,2. 50 NaNs. All mean and sd. calculated on non-NaN values.
				\end{tablenotes}
			\end{footnotesize}
	\end{threeparttable}}
\end{table}

\begin{table}[H]
	\centering
	\noindent\makebox[\textwidth]{%
		\begin{threeparttable}
			\caption{Mean squared error (sd) in estimation and prediction, average Kullback--Leibler divergence, and sensitivity, specificity and precision of variable selection performance, over 50 simulated data sets, $p=200$ and $q=25$. The regression coefficients and precision matrix are estimated by HS-GHS, joint high-dimensional Bayesian variable and covariance selection (BM13), MRCE and CAPME. The best performer in each column is shown in bold.}
			
			\label{stab:3}
			\begin{footnotesize}
				\begin{tabular}{l|ccc|c|ccc|ccc|c|}
					\toprule
					& \multicolumn{11}{|c|}{Simulation 5: $p=200$, $q=25$, $n=100$, Uniform coefficients, Star structure} \\
					& \multicolumn{3}{|c|}{MSE} &
					{Divergence} &
					\multicolumn{3}{|c|}{B support recovery} &
					\multicolumn{3}{|c|}{$\Omega$ support recovery} &
					CPU time \\
					\toprule
					Method & B & $\Omega$ & Prediction & avg KL &
					SEN & SPE & PRC & SEN & SPE & PRC & min. \\
					\toprule
					HS-GHS & \textbf{0.0027} & 0.0341 & \textbf{1.6015} & 10.3918 &
					.9557 & .9975 & \textbf{.9525} & .3856 & .9953 & \textbf{.8523} & 
					1.01e+03 \\
					& (0.0003) & (0.0178) & (0.0686) & (1.4390) &
					(.0115) & (.0008) & (.0145) & (.1277) & (.0041) & (.1130) & \\
					BM13 & 0.0543 & \textbf{0.0150} & 7.1188 & 11.0194 &
					- & - & - & .0011 & .9979 & .0385 $^1$ &
					54.67 \\
					& (0.0006) & (0.0004) & (0.3606) & (0.2732) &
					- & - & - & (.0079) & (.0025) & (.1961) & \\
					MRCE & 0.0865 & 0.0362 & 18.7449 & 32.2416 &
					.0050 & \textbf{1.0000} & .9932 $^2$ & \textbf{.9256} & .0825 & .0607 &
					0.10 \\
					& (0.0003) & (0.0004) & (0.8318) & (0.3412) &
					(.0043) & ($<$.0001) & (.0411) & (.0783) & (.0673) & (.0050) & \\
					CAPME & 0.0096 & 0.0221 & 2.3653 & \textbf{8.2904} &
					\textbf{.9770} & .8098 & .2132 & 0 & \textbf{1} & - $^3$ &
					74.11 \\
					& (0.0009) & (0.0012) & (0.1280) & (0.4024) &
					(.0083) & (.0101) & (.0094) & (0) & (0) & - & \\	
					\toprule
					& \multicolumn{11}{|c|}{Simulation 6: $p=200$, $q=25$, $n=100$, Coefficients=5, AR1 structure} \\
					& \multicolumn{3}{|c|}{MSE} &
					{Divergence} &
					\multicolumn{3}{|c|}{B support recovery} &
					\multicolumn{3}{|c|}{$\Omega$ support recovery} &
					CPU time \\
					\toprule
					Method & B & $\Omega$ & Prediction & avg KL &
					SEN & SPE & PRC & SEN & SPE & PRC & min. \\
					\toprule
					HS-GHS & \textbf{0.0017} & \textbf{0.0400} & \textbf{2.4693} & \textbf{9.5349} &
					\textbf{1} & \textbf{.9986} & \textbf{.9737} & .9817 & .9970 & \textbf{.9672} &
					770.02 \\
					& (0.0002) & (0.0120) & (0.1647) & (1.2234) &
					(0) & (.0006) & (.0108) & (.0281) & (.0039) & (.0396) & \\
					BM13 & 0.7306 & 0.0520 & 80.2711 & 30.2897 &
					- & - & - & 0 & .9898 & 0 &
					103.75 \\
					& (0.0077) & ($<$0.0001) & (4.5903) & (0.2125) &
					- & - & - & (0) & (.0050) & (0) & \\
					MRCE & 1.2326 & 0.1333 & 297.9516 & 66.9764 &
					.0187 & .9903 & .5295 $^4$ & \textbf{.9902} & .0079 & .0799 &
					1.05 \\
					& (0.0159) & (0.2715) & (18.2395) & (18.3250) &
					(.0136) & (.0166) & (.4218) & (.0249) & (.0146) & (.0018) & \\
					CAPME & 0.0202 & 0.0426 & 5.1766 & 14.9741 &
					\textbf{1} & .8070 & .2146 & 0 & \textbf{1} & - $^5$ &
					66.87 \\
					& (0.0022) & (0.0010) & (0.4076) & (0.5903) &
					(0) & (.0097) & (.0083) & (0) & (0) & - & \\
					\bottomrule
				\end{tabular}
				\begin{tablenotes}
					\item 1. 24 NaNs in 50 replicates. 2. 13 NaNs in 50 replicates. 3,5. 50 NaNs. 4. 5 NaNs in 50 replicates. All mean and sd. calculated on non-NaN values.
				\end{tablenotes}
			\end{footnotesize}
	\end{threeparttable}}
\end{table}

\begin{table}[H]
	\centering
	\noindent\makebox[\textwidth]{%
		\begin{threeparttable}
			\caption{Mean squared error (sd) in estimation and prediction, average Kullback--Leibler divergence, and sensitivity, specificity and precision of variable selection performance, over 50 simulated data sets, $p=120$ and $q=50$. The regression coefficients and precision matrix are estimated by HS-GHS, joint high-dimensional Bayesian variable and covariance selection (BM13), MRCE and CAPME. The best performer in each column is shown in bold.}
			
			\label{stab:4}
			\begin{footnotesize}
				\begin{tabular}{l|ccc|c|ccc|ccc|c|}
					\toprule
					& \multicolumn{11}{|c|}{Simulation 3: $p=120$, $q=50$, $n=100$, Uniform coefficients, Star structure} \\
					& \multicolumn{3}{|c|}{MSE} &
					{Divergence} &
					\multicolumn{3}{|c|}{B support recovery} &
					\multicolumn{3}{|c|}{$\Omega$ support recovery} &
					CPU time \\
					\toprule
					Method & B & $\Omega$ & Prediction & avg KL &
					SEN & SPE & PRC & SEN & SPE & PRC & min. \\
					\toprule
					HS-GHS & \textbf{0.0018} & \textbf{0.0036} & \textbf{1.2768} & \textbf{7.3906} &
					.9810 & .9980 & \textbf{.9628} & .4378 & .9995 & \textbf{.9380} &
					2.58e+03 \\
					& (0.0002) & (0.0008) & (0.0454) & (0.7136) &
					(.0069) & (.0006) & (.0114) & (.1445) & (.0007) & (.0980) & \\
					BM13 & 0.0463 & 0.0046 & 3.7041 & 18.4162 &
					- & - & - & .0044 & .9962 & .0190 &
					220.54 \\
					& (0.0006) & (0.0002) & (0.1603) & (0.3480) &
					- & - & - & (.0152) & (.0018) & (.0654) & \\
					MRCE & 0.0856 & 0.0128 & 11.4955 & 47.3913 &
					.0227 & \textbf{.9995} & .8614 & \textbf{.7800} & .2495 & .0156 &
					3.70 \\
					& (0.0021) & (0.0017) & (0.5913) & (1.2672) &
					(.0280) & (.0024) & (.1810) & (.2386) & (.2473) & (.0023) & \\
					CAPME & 0.0072 & 0.0048 & 1.6072 & 9.0566 &
					\textbf{.9893} & .8195 & .2250 & 0 & \textbf{1} & - $^1$ &
					81.38 \\
					& (0.0007) & (0.0006) & (0.0659) & (0.4222) &
					(.0050) & (.0168) & (.0166) & (0) & (0) & - & \\
					\toprule
					& \multicolumn{11}{|c|}{Simulation 4: $p=120$, $q=50$, $n=100$, Coefficients, AR1 structure} \\
					& \multicolumn{3}{|c|}{MSE} &
					{Divergence} &
					\multicolumn{3}{|c|}{B support recovery} &
					\multicolumn{3}{|c|}{$\Omega$ support recovery} &
					CPU time \\
					\toprule
					Method & B & $\Omega$ & Prediction & avg KL &
					SEN & SPE & PRC & SEN & SPE & PRC & min. \\
					\toprule
					HS-GHS & \textbf{0.0014} & \textbf{0.0044} & \textbf{2.3994} & \textbf{7.5001} &
					\textbf{1} & \textbf{.9987} & \textbf{.9757} & .9902 & .9995 & \textbf{.9880} &
					2.56e+03 \\
					& (0.0001) & (0.0010) & (0.0997) & (0.6304) &
					(0) & (.0005) & (.0098) & (.0125) & (.0007) & (.0165) & \\
					BM13 & 0.6005 & 0.0253 & 34.9632 & 54.9328 &
					- & - & - & 0 & .9919 & 0 &
					217.87 \\
					& (0.0084) & ($<$0.0001) & (2.0061) & (0.3322) &
					- & - & - & (0) & (.0019) & (0) & \\
					MRCE & 1.2349 & 0.0259 & 176.9700 & 95.2552 &
					.0207 & .9984 & .4862 $^2$ & \textbf{.9906} & .0105 & .0400 &
					9.76 \\
					& (0.0113) & (0.0003) & (11.3600) & (0.9369) &
					(.0167) & (.0024) & (.2114) & (.0211) & (.0169) & (.0005) & \\
					CAPME & 0.0178 & 0.0188 & 3.8546 & 23.8957 &
					\textbf{1} & .8206 & .2288 & .0184 & \textbf{.9999} & .8491 $^3$ &
					77.02 \\
					& (0.0022) & (0.0011) & (0.2719) & (1.1530) &
					(0) & (.0200) & (.0227) & (.1299) & (.0010) & - & \\				
					\bottomrule
				\end{tabular}
				\begin{tablenotes}
					\item 1. 50 NaNs. 2. 1 NaN in 50 replicates. 3. 49 NaNs in 50 replicates. Mean and sd. calculated on non-NaN values.
				\end{tablenotes}
			\end{footnotesize}
	\end{threeparttable}}
\end{table}

\newpage

\subsection{Assessment of normality assumption for eQTL analysis}
\label{sec:sup_normal}

\begin{figure}[h!]
	\centering
	\adjincludegraphics[width=\textwidth,trim={{0\width} {0.25\height} {0\width} {0.25\height}},clip]{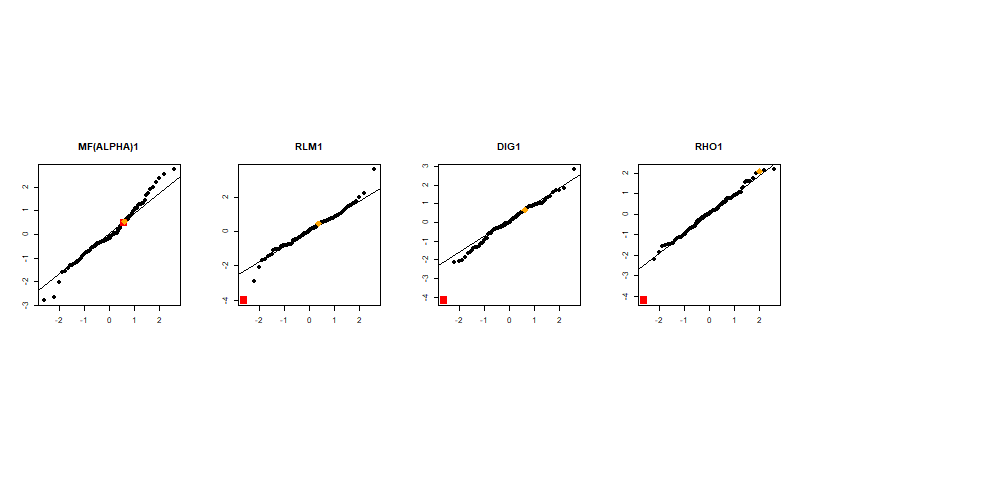}
	\caption{Normal q-q plots of gene expressions. \label{fig:gene}}
\end{figure}

Figure~\ref{fig:gene} shows normal qq-plots of residual gene expression in $54$ MAPK pathway genes. The expressions were regressed on the $172$ markers using lasso regression, and residuals were calculated. Residuals of PKC1, MFA1, SWI6, MFA2 and SSK2 failed univariate Kolmogorov-Smirnov normality test at significance level $0.05$, and these genes were removed from the data set for analysis. Yeast segregants shown in red and orange squares were removed from the data set for analysis.

\begin{figure*}[h!]
	\centering
	\adjincludegraphics[width=\textwidth,trim={{0\width} {0.25\height} {0\width} {0.25\height}},clip]{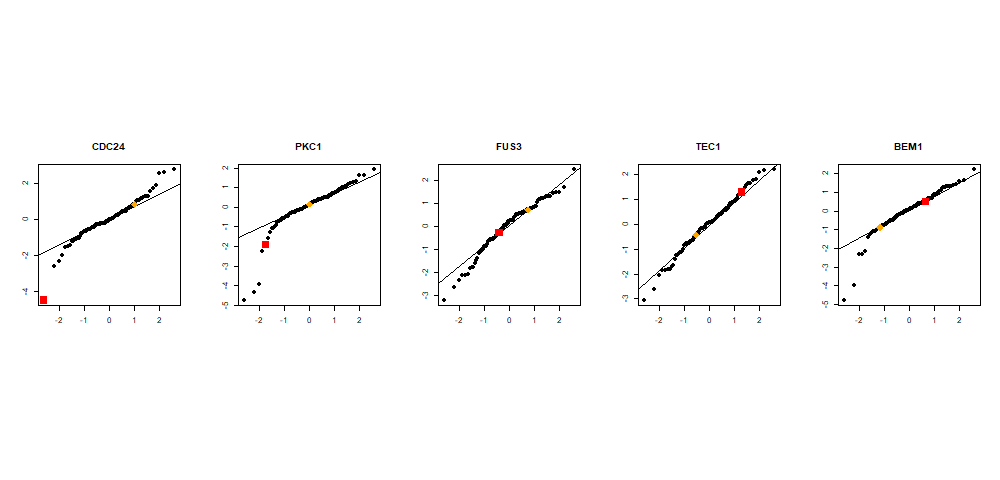}
\end{figure*}
\vspace{-40pt}
\begin{figure*}[h!]
	\centering
	\adjincludegraphics[width=\textwidth,trim={{0\width} {0.25\height} {0\width} {0.25\height}},clip]{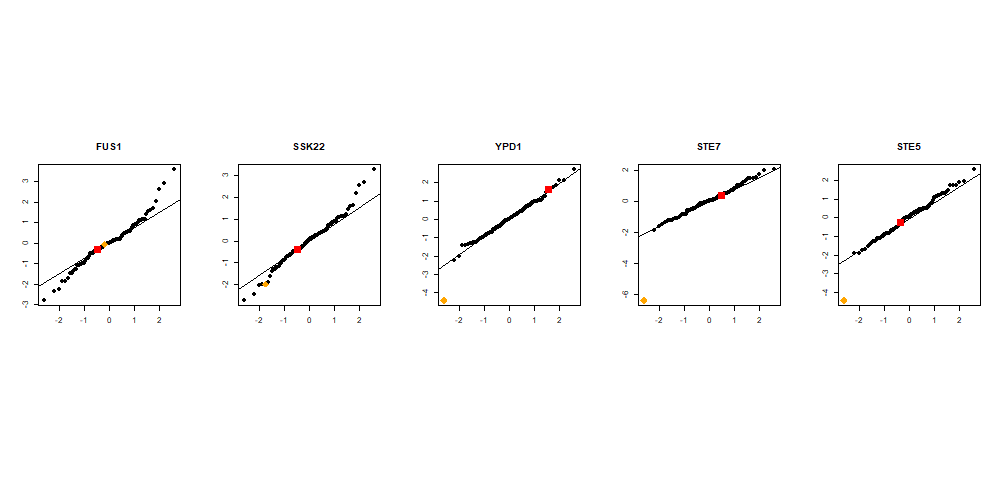}
\end{figure*}
\vspace{-40pt}
\begin{figure*}[h!]
	\centering
	\adjincludegraphics[width=\textwidth,trim={{0\width} {0.25\height} {0\width} {0.25\height}},clip]{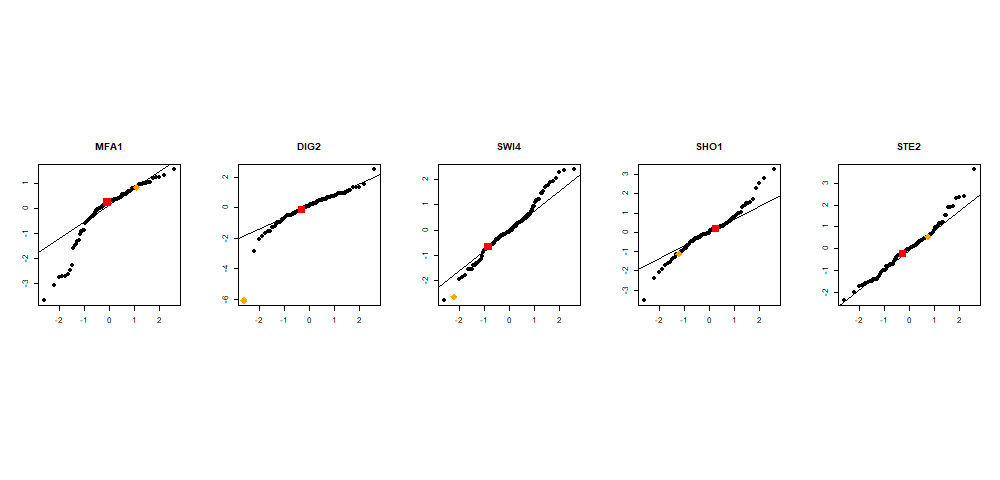}
\end{figure*}
\vspace{-40pt}
\begin{figure*}[h!]
	\centering
	\adjincludegraphics[width=\textwidth,trim={{0\width} {0.25\height} {0\width} {0.25\height}},clip]{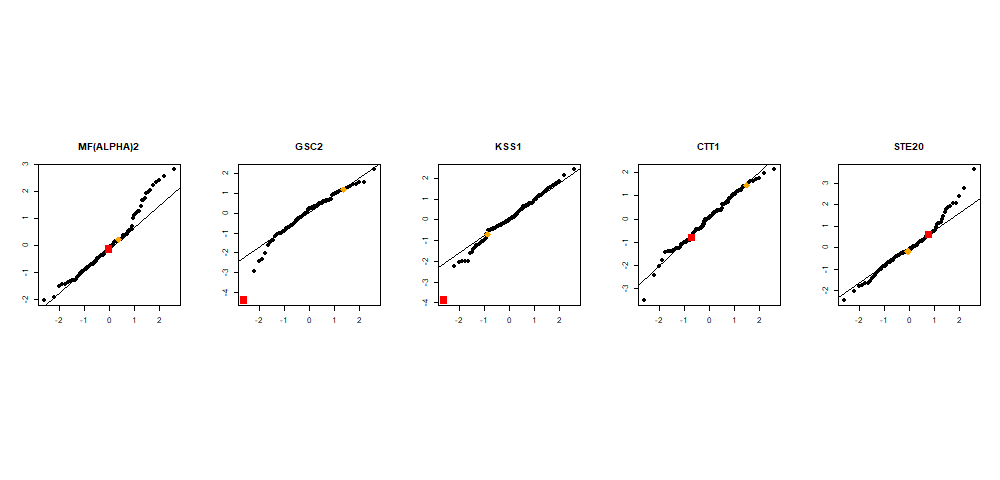}
\end{figure*}
\vspace{-40pt}
\begin{figure*}[h!]
	\centering
	\adjincludegraphics[width=\textwidth,trim={{0\width} {0.25\height} {0\width} {0.25\height}},clip]{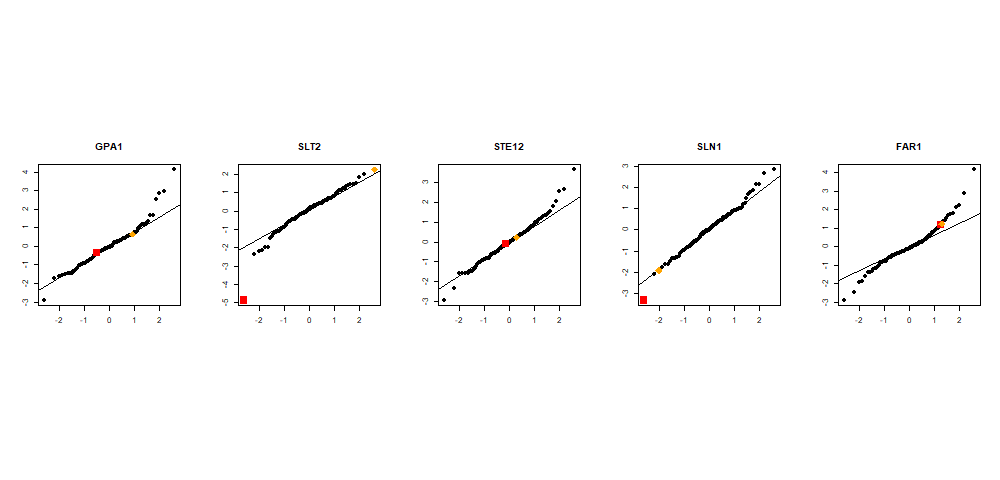}
\end{figure*}
\vspace{-40pt}

\begin{figure*}[h!]
	\centering
	\adjincludegraphics[width=\textwidth,trim={{0\width} {0.25\height} {0\width} {0.25\height}},clip]{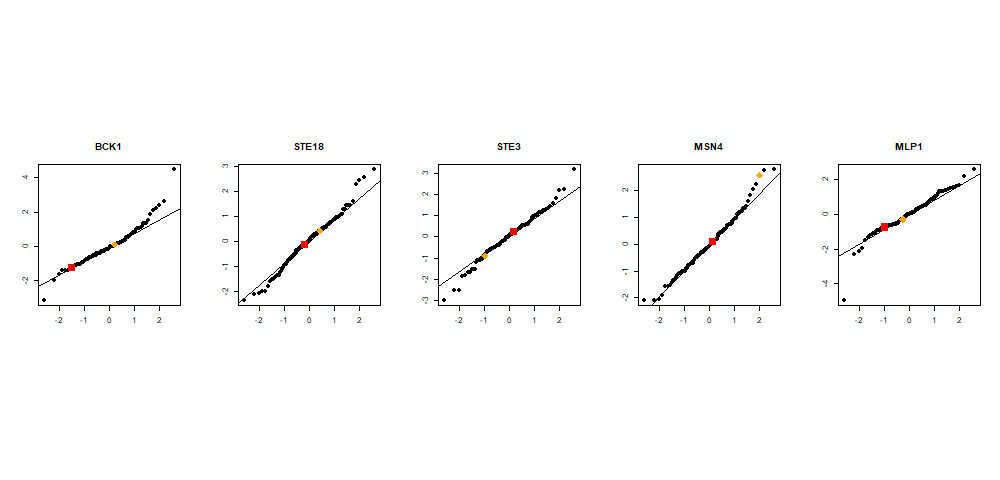}
\end{figure*}
\vspace{-40pt}
\begin{figure*}[h!]
	\centering
	\adjincludegraphics[width=\textwidth,trim={{0\width} {0.25\height} {0\width} {0.25\height}},clip]{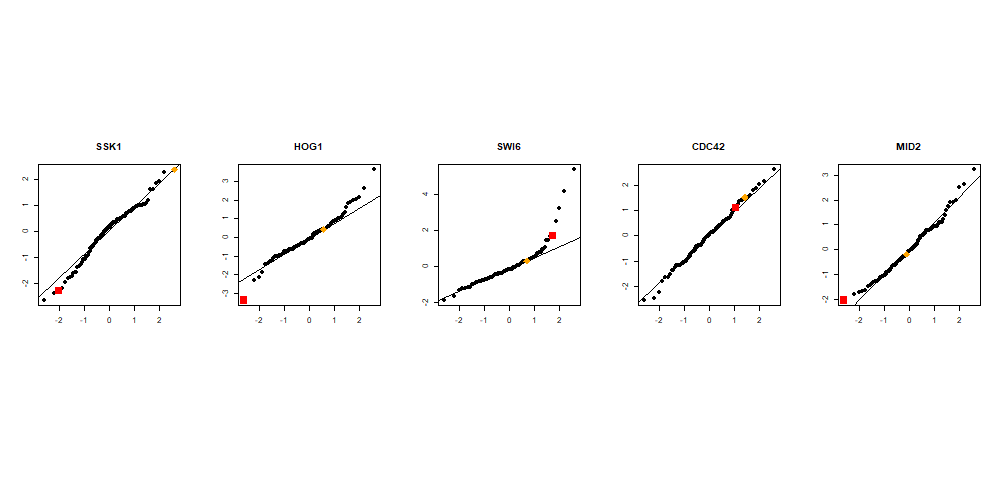}
\end{figure*}
\vspace{-40pt}
\begin{figure*}[h!]
	\centering
	\adjincludegraphics[width=\textwidth,trim={{0\width} {0.25\height} {0\width} {0.25\height}},clip]{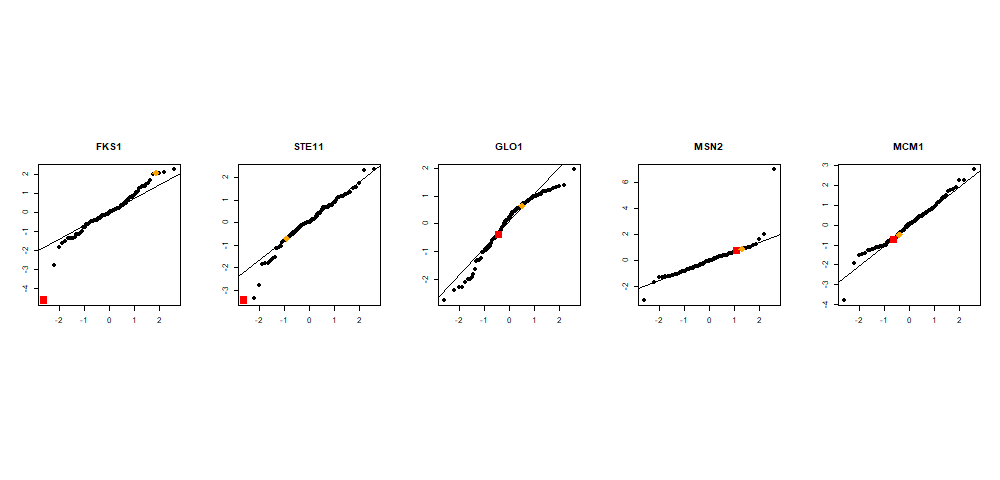}
\end{figure*}
\vspace{-40pt}
\begin{figure*}[h!]
	\centering
	\adjincludegraphics[width=\textwidth,trim={{0\width} {0.25\height} {0\width} {0.25\height}},clip]{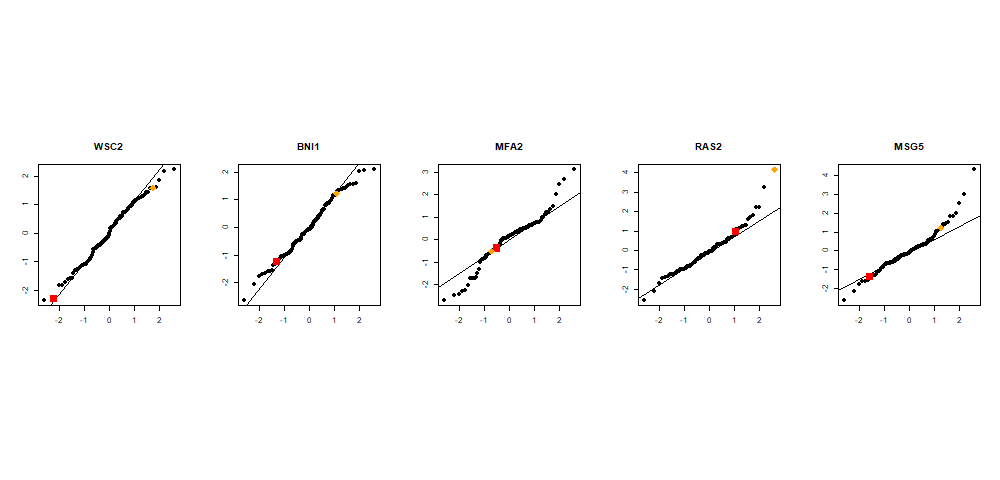}
\end{figure*}
\vspace{-40pt}
\begin{figure*}[h!]
	\centering
	\adjincludegraphics[width=\textwidth,trim={{0\width} {0.25\height} {0\width} {0.25\height}},clip]{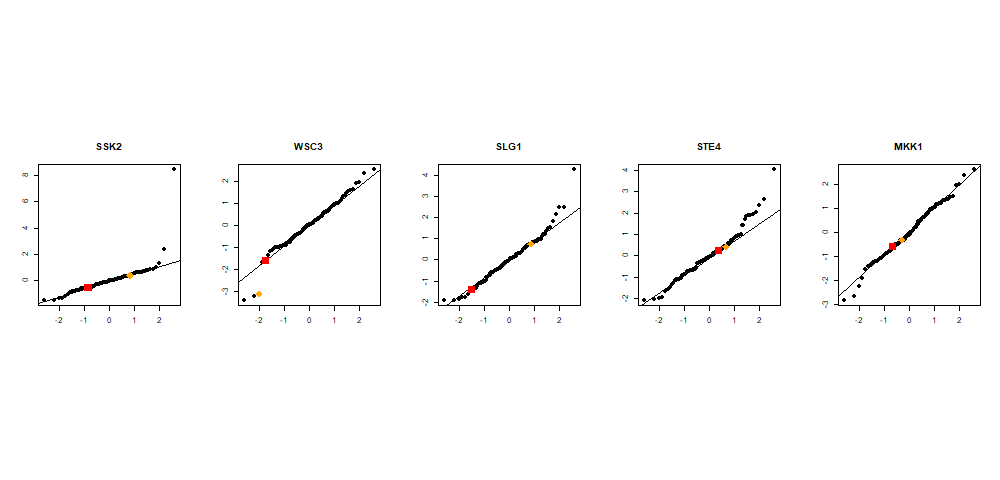}
\end{figure*}
\vspace{-40pt}

\vspace{500pt}
\subsection{Additional eQTL analysis results}
\label{sec:sup_graph}
\begin{figure*}[!h]
	\centering
	\begin{subfigure}[t]{0.48\textwidth}
		\centering
		\adjincludegraphics[width=\textwidth,trim={{0.05\width} {0\height} {0.05\width} {0\height}},clip]{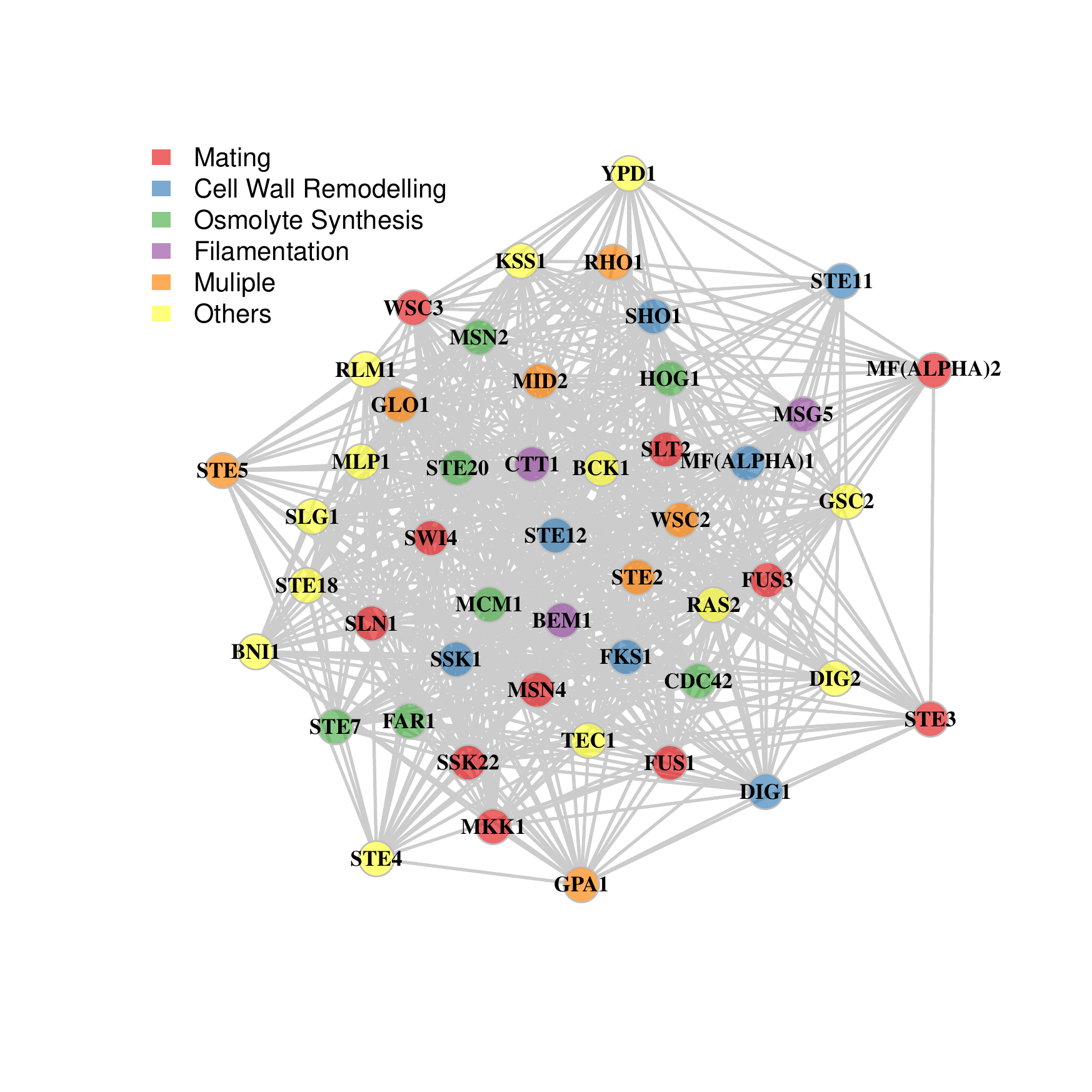}
		\caption[ROC]%
		{{\footnotesize Inferred graph by CAPME estimate}}
		\label{fig:loglik_p120q50}
	\end{subfigure}
	\
	\begin{subfigure}[t]{0.48\textwidth}
		\centering
		\adjincludegraphics[width=\textwidth,trim={{0.05\width} {0\height} {0.05\width} {0\height}},clip]{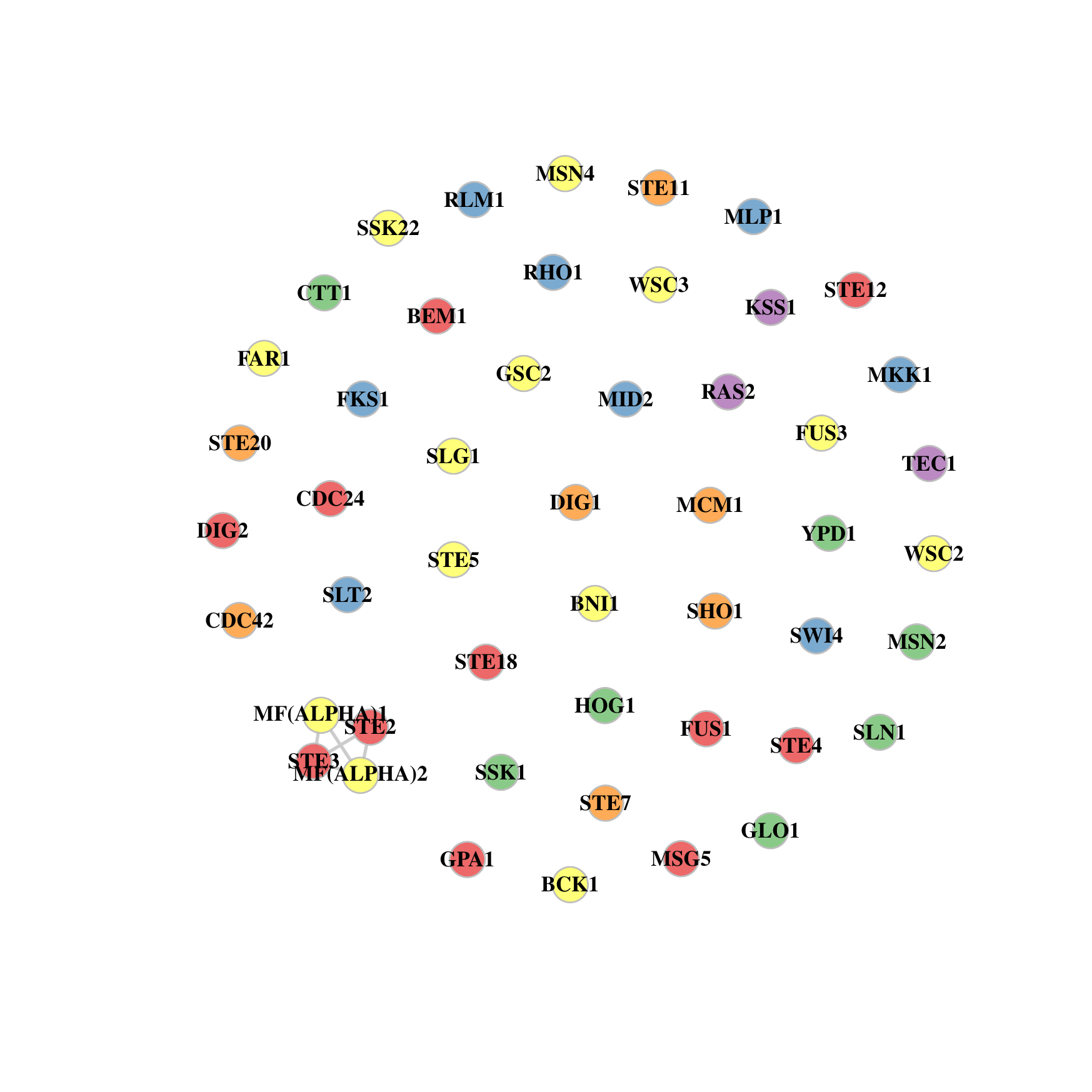}
		\caption[ROC]%
		{{\footnotesize Inferred graph by MRCE estimate}}
		\label{fig:loglik_p200q25}
	\end{subfigure}
	\caption{The inferred graph for the yeast eQTL data, estimated by (a) CAPME, and (b) MRCE.\label{fig:additional}}
\end{figure*}

Figure~\ref{fig:additional} shows the inferred graphs by CAPME and MRCE estimates, complementing the result presented in Figure~\ref{fig:graph} for the proposed HS-GHS estimate.

\end{document}